\documentclass[twocolumn]{aastex63}
\usepackage{amsmath}
\usepackage{graphicx}
\usepackage{natbib}
\usepackage[scaled]{helvet}
\usepackage{epsfig}
\usepackage{url}
\bibpunct{(}{)}{;}{a}{}{,}
\usepackage{textcomp}
\usepackage{mathpazo}
\usepackage{xspace}
\usepackage{hyperref}
\usepackage{apjfonts}
\usepackage{mathrsfs}
\usepackage{verbatim}
\usepackage{threeparttable}

\usepackage{color}
\usepackage[normalem]{ulem}
\usepackage{multirow}

\newcommand{\bjdtdb}{\ensuremath{\rm {BJD_{TDB}}}}
\newcommand{\feh}{\ensuremath{\left[{\rm Fe}/{\rm H}\right]}}

\newcommand{\teff}{\ensuremath{T_{\rm eff}}\xspace}

\newcommand{\ecosw}{\ensuremath{e\cos{\omega_*}}}
\newcommand{\esinw}{\ensuremath{e\sin{\omega_*}}}
\newcommand{\msun}{\ensuremath{\,M_\Sun}}
\newcommand{\rsun}{\ensuremath{\,R_\Sun}}
\newcommand{\lsun}{\ensuremath{\,L_\Sun}}
\newcommand{\mj}{\ensuremath{\,M_{\rm J}}}
\newcommand{\rj}{\ensuremath{\,R_{\rm J}}}

\newcommand{\re}{\ensuremath{\,R_{\rm \Earth}}\xspace}
\newcommand{\me}{\ensuremath{\,M_{\rm \Earth}}\xspace}
\newcommand{\fave}{\langle F \rangle}
\newcommand{\fluxcgs}{10$^9$ erg s$^{-1}$ cm$^{-2}$}

\newcommand{\ktwo}{{\it K2}\xspace}

\newcommand{\be}{\begin{equation}}
\newcommand{\ee}{\end{equation}}

\newcommand{\TESS}{{\it TESS}}

\begin{document}

\title{The K2 \& TESS Synergy III: search and rescue of the lost ephemeris for K2's first planet}

\newcommand{\cfa}{Center for Astrophysics \textbar \ Harvard \& Smithsonian, 60 Garden St, Cambridge, MA 02138, USA}
\newcommand{\msu}{Center for Data Intensive and Time Domain Astronomy, Department of Physics and Astronomy, Michigan State University, East Lansing, MI 48824, USA}
\newcommand{\umich}{Astronomy Department, University of Michigan, 1085 S University Avenue, Ann Arbor, MI 48109, USA}
\newcommand{\utaustin}{Department of Astronomy, The University of Texas at Austin, Austin, TX 78712, USA}
\newcommand{\MIT}{Department of Physics and Kavli Institute for Astrophysics and Space Research, Massachusetts Institute of Technology, Cambridge, MA 02139, USA}
\newcommand{\MITEPS}{Department of Earth, Atmospheric and Planetary Sciences, Massachusetts Institute of Technology,  Cambridge,  MA 02139, USA}
\newcommand{\uflorida}{Department of Astronomy, University of Florida, 211 Bryant Space Science Center, Gainesville, FL, 32611, USA}
\newcommand{\riverside}{Department of Earth and Planetary Sciences, University of California, Riverside, CA 92521, USA}
\newcommand{\usq}{Centre for Astrophysics, University of Southern Queensland, West Street, Toowoomba, QLD 4350, Australia}
\newcommand{\ames}{NASA Ames Research Center, Moffett Field, CA, 94035, USA}
\newcommand{\geneva}{Geneva Observatory, University of Geneva, Chemin des Mailettes 51, 1290 Versoix, Switzerland}
\newcommand{\uw}{Astronomy Department, University of Washington, Seattle, WA 98195 USA}
\newcommand{\warwick}{Department of Physics, University of Warwick, Gibbet Hill Road, Coventry CV4 7AL, UK}
\newcommand{\warwickceh}{Centre for Exoplanets and Habitability, University of Warwick, Gibbet Hill Road, Coventry CV4 7AL, UK}
\newcommand{\princeton}{Department of Astrophysical Sciences, Princeton University, 4 Ivy Lane, Princeton, NJ, 08544, USA}
\newcommand{\liege}{Space Sciences, Technologies and Astrophysics Research (STAR) Institute, Universit\'e de Li\`ege, 19C All\'ee du 6 Ao\^ut, 4000 Li\`ege, Belgium}
\newcommand{\vanderbilt}{Department of Physics and Astronomy, Vanderbilt University, Nashville, TN 37235, USA}
\newcommand{\fisk}{Department of Physics, Fisk University, 1000 17th Avenue North, Nashville, TN 37208, USA}
\newcommand{\columbia}{Department of Astronomy, Columbia University, 550 West 120th Street, New York, NY 10027, USA}
\newcommand{\toronto}{Dunlap Institute for Astronomy and Astrophysics, University of Toronto, Ontario M5S 3H4, Canada}
\newcommand{\unc}{Department of Physics and Astronomy, University of North Carolina at Chapel Hill, Chapel Hill, NC 27599, USA}
\newcommand{\iac}{Instituto de Astrof\'isica de Canarias (IAC), E-38205 La Laguna, Tenerife, Spain}
\newcommand{\lalaguna}{Departamento de Astrof\'isica, Universidad de La Laguna (ULL), E-38206 La Laguna, Tenerife, Spain}
\newcommand{\louisville}{Department of Physics and Astronomy, University of Louisville, Louisville, KY 40292, USA}
\newcommand{\aavso}{American Association of Variable Star Observers, 49 Bay State Road, Cambridge, MA 02138, USA}
\newcommand{\utokyo}{The University of Tokyo, 7-3-1 Hongo, Bunky\={o}, Tokyo 113-8654, Japan}
\newcommand{\naoj}{National Astronomical Observatory of Japan, 2-21-1 Osawa, Mitaka, Tokyo 181-8588, Japan}
\newcommand{\jstpresto}{JST, PRESTO, 7-3-1 Hongo, Bunkyo-ku, Tokyo 113-0033, Japan}
\newcommand{\astrobiojapan}{Astrobiology Center, 2-21-1 Osawa, Mitaka, Tokyo 181-8588, Japan}
\newcommand{\ctio}{Cerro Tololo Inter-American Observatory, Casilla 603, La Serena, Chile}
\newcommand{\noirlab}{NOIRLab/Cerro Tololo Inter-American Observatory, Casilla 603, La Serena, Chile}
\newcommand{\nexsci}{Caltech IPAC -- NASA Exoplanet Science Institute 1200 E. California Ave, Pasadena, CA 91125, USA}
\newcommand{\ucsc}{Department of Astronomy and Astrophysics, University of
California, Santa Cruz, CA 95064, USA}
\newcommand{\gsfc}{Exoplanets and Stellar Astrophysics Laboratory, Code 667, NASA Goddard Space Flight Center, Greenbelt, MD 20771, USA}
\newcommand{\sgtinc}{SGT, Inc./NASA AMES Research Center, Mailstop 269-3, Bldg T35C, P.O. Box 1, Moffett Field, CA 94035, USA}
\newcommand{\chile}{Center of Astro-Engineering UC, Pontificia Universidad Cat\'olica de Chile, Av. Vicu\~{n}a Mackenna 4860, 7820436 Macul, Santiago, Chile}
\newcommand{\Pontificia}{Facultad de Ingeniería y Ciencias, Universidad Adolfo Ib\'a\~nez, Av. Diagonal las Torres 2640, Pe\~nalol\'en, Santiago, Chile}
\newcommand{\Millennium}{Millennium Institute for Astrophysics, Chile}
\newcommand{\maxplank}{Max-Planck-Institut f\"ur Astronomie, K\"onigstuhl 17, Heidelberg 69117, Germany}
\newcommand{\utdallas}{Department of Physics, The University of Texas at Dallas, 800 West
Campbell Road, Richardson, TX 75080-3021 USA}
\newcommand{\MauryLewin}{Maury Lewin Astronomical Observatory, Glendora, CA 91741, USA}
\newcommand{\umbc}{University of Maryland, Baltimore County, 1000 Hilltop Circle, Baltimore, MD 21250, USA}
\newcommand{\osu}{Department of Astronomy, The Ohio State University, 140 West 18th Avenue, Columbus, OH 43210, USA}
\newcommand{\MITAA}{Department of Aeronautics and Astronautics, MIT, 77 Massachusetts Avenue, Cambridge, MA 02139, USA}
\newcommand{\openu}{School of Physical Sciences, The Open University, Milton Keynes MK7 6AA, UK}
\newcommand{\swarthmore}{Department of Physics and Astronomy, Swarthmore College, Swarthmore, PA 19081, USA}
\newcommand{\seti}{SETI Institute, Mountain View, CA 94043, USA}
\newcommand{\lehigh}{Department of Physics, Lehigh University, 16 Memorial Drive East, Bethlehem, PA 18015, USA}
\newcommand{\utah}{Department of Physics and Astronomy, University of Utah, 115 South 1400 East, Salt Lake City, UT 84112, USA}
\newcommand{\USNA}{Department of Physics, United States Naval Academy, 572C Holloway Rd., Annapolis, MD 21402, USA}
\newcommand{\DTM}{Department of Terrestrial Magnetism, Carnegie Institution for Science, 5241 Broad Branch Road, NW, Washington, DC 20015, USA}
\newcommand{\UPenn}{The University of Pennsylvania, Department of Physics and Astronomy, Philadelphia, PA, 19104, USA}
\newcommand{\montana}{Department of Physics and Astronomy, University of Montana, 32 Campus Drive, No. 1080, Missoula, MT 59812 USA}
\newcommand{\psu}{Department of Astronomy \& Astrophysics, The Pennsylvania State University, 525 Davey Lab, University Park, PA 16802, USA}
\newcommand{\psust}{Center for Exoplanets and Habitable Worlds, The Pennsylvania State University, 525 Davey Lab, University Park, PA 16802, USA}
\newcommand{\Kutztown}{Department of Physical Sciences, Kutztown University, Kutztown, PA 19530, USA}
\newcommand{\udel}{Department of Physics \& Astronomy, University of Delaware, Newark, DE 19716, USA}
\newcommand{\Westminster}{Department of Physics, Westminster College, New Wilmington, PA 16172}
\newcommand{\steward}{Department of Astronomy and Steward Observatory, University of Arizona, Tucson, AZ 85721, USA}
\newcommand{\saao}{South African Astronomical Observatory, PO Box 9, Observatory, 7935, Cape Town, South Africa}
\newcommand{\salt}{Southern African Large Telescope, PO Box 9, Observatory, 7935, Cape Town, South Africa}
\newcommand{\ssl}{Societ\`{a} Astronomica Lunae, Italy}
\newcommand{\spot}{Spot Observatory, Nashville, TN 37206, USA}
\newcommand{\txamGP}{George P.\ and Cynthia Woods Mitchell Institute for Fundamental Physics and Astronomy, Texas A\&M University, College Station, TX77843 USA}
\newcommand{\txam}{Department of Physics and Astronomy, Texas A\&M university, College Station, TX 77843 USA}
\newcommand{\txammul}{Munnerlyn Astronomical Instrumentation Laboratory, Department of Physics \& Astronomy, Texas A\&M university, College Station, TX 77843 USA}
\newcommand{\wellesley}{Department of Astronomy, Wellesley College, Wellesley, MA 02481, USA}
\newcommand{\Wesleyan}{Astronomy Department and Van Vleck Observatory, Wesleyan University, Middletown, CT 06459, USA}
\newcommand{\inaf}{INAF -- Osservatorio Astronomico di Padova, Vicolo dell'Osservatorio 5, I-35122, Padova, Italy}
\newcommand{\byu}{Department of Physics and Astronomy, Brigham Young University, Provo, UT 84602, USA}
\newcommand{\Hazelwood}{Hazelwood Observatory, Churchill, Victoria, Australia}
\newcommand{\pest}{Perth Exoplanet Survey Telescope}
\newcommand{\Winer}{Winer Observatory, PO Box 797, Sonoita, AZ 85637, USA}
\newcommand{\icpo}{Ivan Curtis Private Observatory}
\newcommand{\elsauce}{El Sauce Observatory, Chile}
\newcommand{\crow}{Atalaia Group \& CROW Observatory, Portalegre, Portugal}
\newcommand{\dfus}{Dipartimento di Fisica ``E.R.Caianiello'', Universit\`a di Salerno, Via Giovanni Paolo II 132, Fisciano 84084, Italy}
\newcommand{\indfn}{Istituto Nazionale di Fisica Nucleare, Napoli, Italy}
\newcommand{\sotes}{Gabriel Murawski Private Observatory (SOTES)}
\newcommand{\lco}{Las Cumbres Observatory Global Telescope, 6740 Cortona Dr., Suite 102, Goleta, CA 93111, USA}
\newcommand{\ucsb}{Department of Physics, University of California, Santa Barbara, CA 93106-9530, USA}
\newcommand{\yale}{Department of Astronomy, Yale University, 52 Hillhouse Avenue, New Haven, CT 06511, USA}
\newcommand{\eso}{European Southern Observatory, Alonso de C\'ordova 3107, Vitacura, Casilla 19001, Santiago, Chile}
\newcommand{\stsci}{Space Telescope Science Institute, Baltimore, MD 21218, USA}
\newcommand{\keele}{Astrophysics Group, Keele University, Staffordshire ST5 5BG, UK}
\newcommand{\gsfcsellers}{GSFC Sellers Exoplanet Environments Collaboration, NASA Goddard Space Flight Center, Greenbelt, MD 20771 }
\newcommand{\usno}{U.S. Naval Observatory, Washington, DC 20392, USA}
\newcommand{\kansas}{Department of Physics and Astronomy, University of Kansas, Lawrence, KS 66045, USA}
\newcommand{\gmu}{George Mason University, 4400 University Drive MS 3F3, Fairfax, VA 22030, USA}
\newcommand{\unsw}{Exoplanetary Science at UNSW, School of Physics, UNSW Sydney, NSW 2052, Australia}
\newcommand{\sifa}{School of Physics, Sydney Institute for Astronomy (SIfA), The University of Sydney, NSW 2006, Australia}
\newcommand{\nanjing}{School of Astronomy and Space Science, Key Laboratory of Modern Astronomy and Astrophysics in Ministry of Education, Nanjing University, Nanjing 210046, Jiangsu, China}
\newcommand{\berkely}{Department of Astronomy, University of California Berkeley, Berkeley, CA 94720-3411, USA}
\newcommand{\bhicfa}{Black Hole Initiative at Harvard University, 20 Garden Street, Cambridge, MA 02138, USA}
\newcommand{\Silesian}{Department of Electronics, Electronical Engineering and Microelectronics, Silesian University of Techhnology Akademicka 16, 44-100 Gliwice, Poland}
\newcommand{\Patashnick}{Patashnick Voorheesville Observatory, Voorheesville, NY 12186, USA}
\newcommand{\austincollege}{Physics Department, Austin College, 900 North Grand Avenue, Sherman TX 75090, USA}
\newcommand{\Tsinghua}{Department of Astronomy, Tsinghua University, Beijing 100084, China}
\newcommand{\Tsinghuaschool}{Tsinghua International School, Beijing 100084, China}
\newcommand{\chinaNAO}{National Astronomical Observatories, Chinese Academy of Sciences, 20A Datun Road, Chaoyang District, Beijing 100012, China}
\newcommand{\Tautenburg}{Th{\"u}ringer Landessternwarte Tautenburg, Sternwarte 5, 07778 Tautenburg, Germany}
\newcommand{\brierfield}{Brierfield Observatory, New South Wales, Australia}
\newcommand{\Indiana}{Indiana University Department of Astronomy, SW319, 727 E 3rd Street, Bloomington, IN 47405 USA}
\newcommand{\wisconsin}{Department of Astronomy, University of Wisconsin-Madison, Madison, WI 53706, USA}
\newcommand{\protologic}{Proto-Logic Consulting LLC, Washington, DC 20009, USA}
\newcommand{\ASTRAVEO}{ASTRAVEO LLC, PO Box 1668, MA 01931}
\newcommand{\TJHS}{Thomas Jefferson High School, 6560 Braddock Rd, Alexandria, VA 22312 USA}
\newcommand{\ucatchile}{Instituto de Astrof\'isica, Facultad de F\'isica, Pontificia Universidad Cat\'olica de Chile}
\newcommand{\lasa}{Liberal Arts and Science Academy, Austin, Texas 78724, USA}
\newcommand{\gemini}{Gemini Observatory/NSF’s NOIRLab, 670 N. A’ohoku Place, Hilo, HI, 96720, USA}
\newcommand{\umd}{Department of Astronomy, University of Maryland, College Park, College Park, MD}
\newcommand{\ucscchile}{Departamento de Matem\'atica y F\i'sica Aplicadas, Universidad Cat\'olica de la Sant\'isima Concepci\'on, Alonso de Rivera 2850, Concepci\'on, Chile}
\newcommand{\jpl}{NASA Jet Propulsion Laboratory, Pasadena, CA, USA}
\newcommand{\asp}{Astronomical Science Program, Graduate University for Advanced Studies, SOKENDAI, 2-21-1, Osawa, Mitaka, Tokyo, 181-8588, Japan}
\newcommand{\iauk}{Institute of Astronomy, University of Cambridge, Madingley
  Road, Cambridge, CB3 0HA, United Kingdom}
\newcommand{\oxford}{Department of Physics, University of Oxford, Oxford OX13RH, United Kingdom}


\newcommand{\eberly}{\altaffiliation{Eberly Research Fellow}}
\newcommand{\torres}{\altaffiliation{Juan Carlos Torres Fellow}}
\newcommand{\sagan}{\altaffiliation{NASA Sagan Fellow}}
\newcommand{\bernoulli}{\altaffiliation{Bernoulli fellow}}
\newcommand{\gruber}{\altaffiliation{Gruber fellow}}
\newcommand{\kavli}{\altaffiliation{Kavli Fellow}}
\newcommand{\peg}{\altaffiliation{51 Pegasi b Fellow}}
\newcommand{\pappalardo}{\altaffiliation{Pappalardo Fellow}}
\newcommand{\hubble}{\altaffiliation{NASA Hubble Fellow}}
\newcommand{\nsf}{\altaffiliation{National Science Foundation Graduate Research Fellow}}
\newcommand{\quadfel}{\altaffiliation{Quad Fellow}}
\newcommand{\mitfel}{\altaffiliation{MIT Collamore-Rogers Fellow}}

\correspondingauthor{Erica Thygesen} 
\email{thygesen@msu.edu}

\author[0000-0002-9165-6245]{Erica Thygesen} 
\quadfel
\affiliation{\msu}


\author[0000-0001-8812-0565]{Joseph E. Rodriguez} 
\affiliation{\msu}

\author[0000-0001-8812-0565]{Zo\"e L. de Beurs}
\nsf
\mitfel
\affiliation{\MITEPS}

\author[0000-0001-7246-5438]{Andrew Vanderburg} 
\affiliation{\MIT}

\author[0000-0002-4881-3620]{John H. Livingston}
\affiliation{\astrobiojapan}
\affiliation{\naoj}
\affiliation{\asp}

\author{Jonathon Irwin}
\affiliation{\iauk}

\author[0000-0002-8400-1646]{Alexander Venner}
\affiliation{\usq}

\author[0000-0002-2207-0750]{Michael Cretignier}
\affiliation{\oxford}

\author[0000-0001-6588-9574]{Karen A.\ Collins}
\affiliation{\cfa}

\author[0000-0001-6637-5401]{Allyson Bieryla} 
\affiliation{\cfa}

\author[0000-0002-9003-484X]{David Charbonneau}
\affiliation{\cfa}

\author[0000-0002-1835-1891]{Ian J.\ M.\ Crossfield}
\affiliation{\kansas}

\author[0000-0002-9332-2011]{Xavier Dumusque}
\affiliation{\geneva}

\author[0000-0003-0497-2651]{John Kielkopf}
\affiliation{\louisville}

\author[0000-0001-9911-7388]{David W. Latham} 
\affiliation{\cfa}

\author{Michael Werner}
\affiliation{\jpl}




















\shorttitle{The \ktwo\ \& \TESS\ Synergy III}
\shortauthors{Thygesen et al.}

\begin{abstract}


K2-2 b/HIP 116454 b, the first exoplanet discovery by K2 during its Two-Wheeled Concept Engineering Test, is a sub-Neptune ($2.5\pm0.1~R_\oplus$, $9.7 \pm 1.2~M_{\oplus}$) orbiting a relatively bright (K$_S$ = 8.03) K-dwarf on a 9.1 day period. Unfortunately, due to a spurious follow-up transit detection and ephemeris degradation, the transit ephemeris for this planet was lost. In this work, we recover and refine the transit ephemeris for K2-2 b, showing a $\sim$40$\sigma$ discrepancy from the discovery results. To accurately measure the transit ephemeris and update the parameters of the system, we jointly fit space-based photometric observations from NASA's K2, TESS, and Spitzer missions with new photometric observations from the ground, as well as radial velocities from HARPS-N that are corrected for stellar activity using a new modeling technique. Ephemerides becoming lost or significantly degraded, as is the case for most transiting planets, highlights the importance of systematically updating transit ephemerides with upcoming large efforts expected to characterize hundreds of exoplanet atmospheres. K2-2 b sits at the high-mass peak of the known radius valley for sub-Neptunes, and is now well-suited for transmission spectroscopy with current and future facilities. Our updated transit ephemeris will ensure no more than a 13-minute uncertainty through 2030.  


\end{abstract}

\section{Introduction}
\label{sec:intro}

In the era of cutting-edge atmospheric characterization of transiting exoplanets, precise and accurate ephemerides are crucial for efficiently scheduling these expensive observations. However, over 80$\%$ of transiting exoplanets will have uncertainties on their future transit times greater than 30 minutes by the end of the decade (see \citealt{Thygesen:2023}), rendering these systems extremely challenging to observe with JWST \citep{Gardner:2006,Beichman:2020}, major upcoming facilities such as the Atmospheric Remote-sensing Infrared Exoplanet Large-survey (ARIEL; \citealt{Tinetti-ARIEL:2018,Tinetti-ARIEL:2021}), and 30m class telescopes like the Thirty Meter Telescope (TMT; \citealt{sanders:2013TMT}), Giant Magellan Telescope \citep{Johns:2012GMT}, and the 39 m European Southern Observatory Extremely Large Telescope (ELT; \citealt{Udry-ELT:2014}). This problem can be solved by observing new transits of these planets with current facilities. Fortunately, NASA's Transiting Exoplanet Survey Satellite (TESS) mission \citep{Ricker:2015} is observing the entire sky, providing a valuable opportunity to refine the transit ephemeris for most known planets. 

After a successful 4-year nominal mission, discovering thousands of exoplanets, the Kepler mission \citep{Borucki:2010} was repurposed due to a mechanical issue. Using the solar pressure to stabilize pointing of the Kepler spacecraft, the K2 mission was able to survey the ecliptic plane, finding hundreds of exciting new systems that are well-suited for detailed characterization \citep{Howell:2012,Vanderburg:2016b,Zink:2021,Kruse:2019,Pope:2016,Livingston:2018-K2C5-8,Crossfield:2016,Dattilo:2019}. The K2 mission ended in 2019, with many of its newly-detected planets never being reobserved since their discovery campaign(s). The K2 \& TESS Synergy project is an effort to provide the community with updated and accurate transit times and system parameters for exoplanets originally discovered by the K2 mission that have been recently observed by TESS \citep{Ricker:2015}. Following a successful pilot study \citep{Ikwut-Ukwa-SynergyI:2020}, the second paper in this series revisited 26 K2 single-planet systems that TESS reobserved during its prime mission \citep{Thygesen:2023}. This work improved the average ephemeris uncertainties by multiple orders of magnitude due to the addition of new TESS transits. Additionally, we identified systems where the original ephemeris has been completely lost (See K2-260; \citealt{Thygesen:2023}), which is similar to this work on K2-2 b, K2's first exoplanet discovery.


K2-2 b was identified during the Two-Wheeled Concept Engineering Test (campaign 0) of the K2 mission. K2-2 b is a sub-Neptune ($2.5\pm0.1~R_\oplus$, $9.7 \pm 1.2~M_{\oplus}$) on a 9.1-day orbit around a bright ($V=10.2,~J=8.6$, HIP 116454) K-dwarf \citep{Vanderburg:2015-K2-2}. At discovery, a single clear transit was detected in the K2 observations, along with a marginal ($\sim$3$\sigma$) detection from the Microvariablity and Oscillations of Stars (MOST) Space Telescope \citep{Walker:2003-MOST}. Follow-up observations were scheduled with Spitzer (P.I. Werner, AOR 57185280) and the Hubble Space Telescope (P.I. Bourrier, proposal I.D. 15127), however, the transit was not seen during the predicted window from the discovery ephemeris. It was then determined that the MOST transit was likely not a real transit of K2-2 b, having skewed the period enough to cause subsequent transits to be missed. 

In this work, we combine the discovery observations from \cite{Vanderburg:2015-K2-2} with new observations from NASA's TESS mission, follow up ground-based photometry, and improved radial velocities to accurately measure the ephemeris of K2-2 b for the first time, proving the original detection from MOST to be a false positive. In Section \ref{sec:obs} we describe the observations used and the relevant reduction and analysis methods, including the reanalysis of radial velocities from the High Accuracy Radial Velocity Planet Searcher-North (HARPS-N; \citet{Consentino:2012}) on the 3.58m Telescopio Nazionale Galileo at the Roque de los Muchachos Observatory. Section \ref{sec:GlobalFit} outlines the methodology used in running the \texttt{EXOFASTv2} global fit of all observations and archival information. We present our results and discuss the importance of ephemeris refinement in the context of future characterization of K2-2 b in Section \ref{sec:discussion}.



\section{Observations and Archival Data} \label{sec:obs}

\begin{figure}
    \centering
    \includegraphics[width=\linewidth]{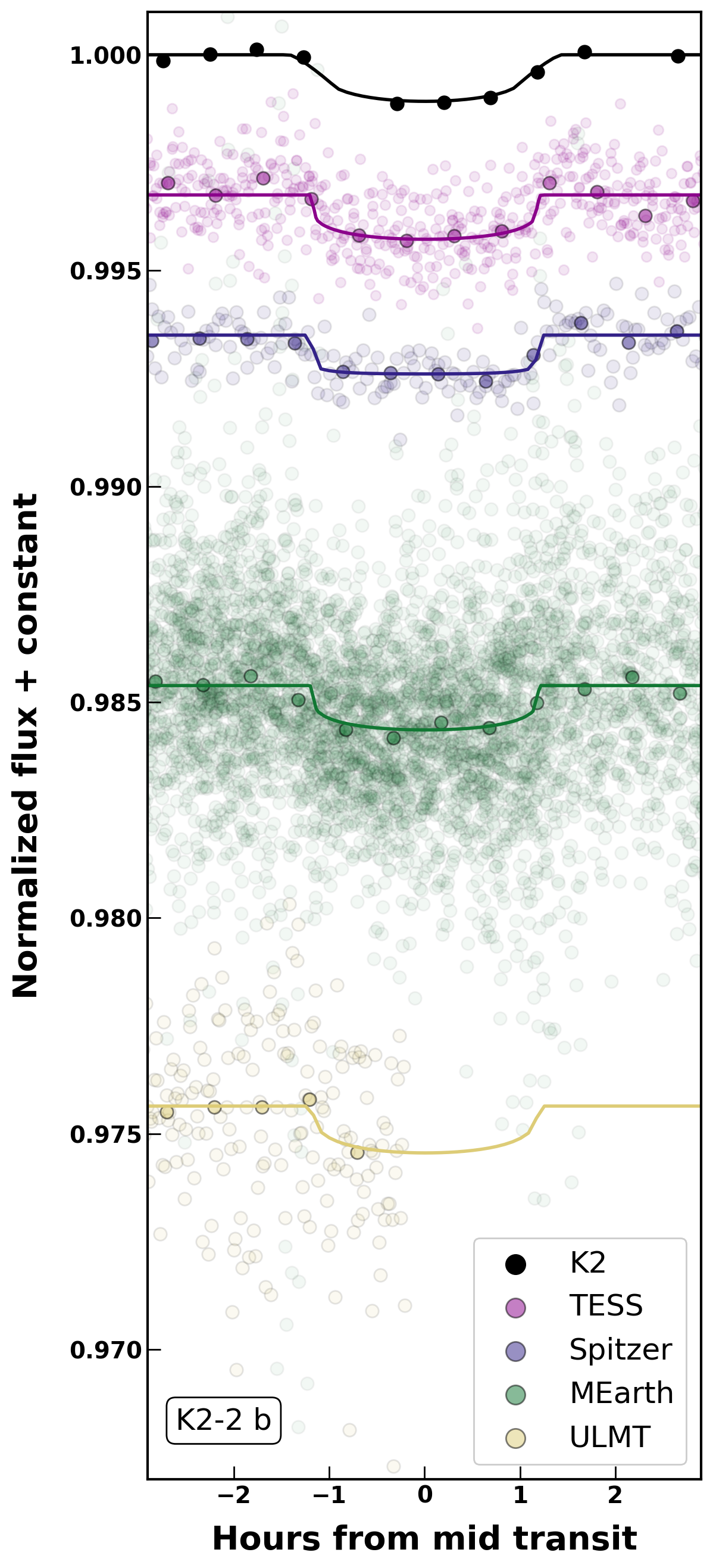}
    \caption{The discovery and follow up phase-folded transits of K2-2 b used in the EXOFASTv2 (see Section \ref{sec:GlobalFit}) analysis. The observations from K2 (black), TESS (purple), Spitzer (blue), MEarth (green), and ULMT (yellow) are shown in open colored circles with the solid colored line representing the EXOFASTv2 model for that dataset. The closed colored circles represent 30-minute bins. East transit is offset by a constant for clarity.} 
    \label{fig:transits}
\end{figure}

The discovery analysis for K2-2 b included a 47 day long light curve from MOST \citep{Walker:2003-MOST}, which was thought to contain a marginal $\sim 3\sigma$ detection of the transit, but future follow up attempts to reobserve the transit with Spitzer and HST showed no transit during or near the predicted window. This ultimately led to the idea that the MOST observations were not reliably constraining the transit ephemeris. While it is not clear why this happened, it is possible that Gaussian noise or satellite systematics caused an already marginal detection to be anchored to a different time of transit. Our new observations from MEarth, ULMT, Spitzer and TESS (Figure \ref{fig:transits}) confirm this hypothesis. In the near decade since its discovery, a variety of follow up observations have been conducted to better characterize the K2-2 system and to recover the transit ephemeris. In the following sections, we describe the new and archival observations used in our analysis. The magnitudes and literature values for K2-2 are listed in Table \ref{tab:literature}, and the photometric data sets we used are outlined in Table \ref{tab:obs}.



\subsection{Ground-based archival imaging}
At the discovery of of K2-2 b, \citet{Vanderburg:2015-K2-2} used multiple archival from the National Geographic Society–Palomar Observatory Sky Survey (POSS-I, \citealt{vanLeeuwen:2007}) and Sloan Digital Sky Survey (SDSS, \citealt{Abazajian:2009}), and newly acquired images from Robo-AO on Palomar \citep{Baranec:2014, Law:2014} and Natural Guide Star Adaptive Optics (NGSAO) system on Keck to rule out nearby close companions that might be contaminating the K2 aperture. A nearby white dwarf with a separation of around 8$\arcsec$ was identified to share a similar proper motion to K2-2, suggesting that they exist in a gravitationally bound system (this is discussed more in Section \ref{sec:futurework}). The white dwarf is within the K2 aperture, but is 6-7 magnitudes dimmer than K2-2, which would not affect the final transit depth of K2-2 b. No other nearby companions were found to a 7$\sigma$ significance in the $H$ band to the limits of 3.0 mag at $0\arcsec.1$ separation, 9.2 mag at $1\arcsec.0$ and 12.7 mag at $5\arcsec.0$.   


\subsection{{\it K2} Photometry}
\label{sec:k2}

A single transit of K2-2 b was observed at 30-minute cadence during the Kepler Two-Wheel Concept Engineering Test during February 2014. Due to the loss of two of the four reaction wheels on the spacecraft, significant systematics were introduced to the light curves of the K2 mission. We corrected for these using the methods described in \citet{Vanderburg:2014} and \citet{Vanderburg:2016b}, which utilize a series of 20 apertures to extract raw light curves used to perform the corrections. Short timescale variations in each of these light curves are correlated with the roll angle of the spacecraft, with the latter being subtracted from the light curves. This process is repeated iteratively until the light curve is free of any variations associated with the roll of the spacecraft. The most precise light curve out of the 20 following the corrections is selected for final analysis. We performed further corrections by fitting the transit and correcting for the systematics and any low-frequency stellar variability, prior to the global fit. 

\subsection{MEarth}

MEarth was used to initially recover the transit of K2-2 b and constrain the ephemeris, observing multiple partial and full transits. MEarth consists of 16 separate 0.4 m telescopes using custom 715 nm longpass filters designed to find Earth-sized planets around M dwarfs \citep{Nutzman:2008,mearth:2015}. Telescopes 1-8 are a part of the MEarth-North Observatory at Fred Lawrence Whipple Observatory (FLWO) on Mount Hopkins, Arizona, while the other eight telescopes (numbered as 11-18) are part of the MEarth-South Observatory located at Cerro Tololo Inter-American Observatory (CTIO) on Cerro Tololo, Chile. K2-2 was observed using a subset of four telescopes from each observatory (see Table \ref{tab:obs}) with 1 minute cadence on UT 2016 September 21 and 30, and UT 2016 October 09. Light curves from MEarth are automatically extracted through a pipeline (see \citealt{Irwin:2007,Berta:2011}) that calibrates the images using flat fields, dark current frames and bias exposures. We combined the light curves across multiple nights for each telescope, so within the global fit the variance can be determined independently for each instrument. We sliced the light curves such that we only included one full transit duration before and after the transit, and detrended against airmass in the global fit. While the original observations also included telescopes 4, 5 and 8, we did not use these in our analysis as the light curves did not contain full transits and would not contribute significant value to the global fit. The transit was also missed during the night of UT 2016 September 11 due to the incorrect ephemeris.

These observations were the first use of the defocus observing mode of
MEarth for transit follow-up, and served as the prototype for a large
number of observations of TESS objects of interest done in later
years.  Here we describe the modifications made to the system to
implement this mode.  Prior to implementation of defocus, MEarth
observations of bright stars were limited by scintillation noise due
to the short maximum exposure times possible before detector
saturation, combined with high overheads (approximately 15s, most of
which was consumed by CCD readout and download over USB2 connection to
the host computer), resulting in a low duty cycle.  For scintillation
limited observations of events of fixed duration such as transits, the
overall transit-averaged photometric noise is determined by the duty
cycle (e.g. \citealt{1967AJ.....72Q.328Y}) so the goal of implementing
defocus was to improve this by substantially lengthening the exposure
times possible before saturation.

The scheduling and telescope control software were modified to allow
each observation request to specify defocus as half flux diameter
(HFD), in pixels.  For these first observations of K2-2, we used HFD =
6.0 pixels, where the pixel scales are 0.76 arcsec/pix for
MEarth-North and 0.84 arcsec/pix for MEarth-South.  The telescope
focus was offset by the scheduler prior to commencing observations of
each target by the appropriate number of focus encoder counts, where
the scaling factor was determined from the calibration curve of HFD
versus focus encoder counts used by the standard automatic focus
routine (normally used for focusing the telescope at the start of the
night).

MEarth did not have autoguiders, and guiding to stabilize the target
star position on the detector (vital for precise transit work) had to
be done using the science exposures themselves, which were 36s for K2-2.
The standard MEarth target acquisition and guiding system for normal
in-focus images consisted of astrometric analysis of the images after
readout to determine their center in celestial coordinates, followed
by offsetting of the telescope to center the target based on its
calculated position.  Target acquisition was done by applying the full
offset, and guiding by passing these measurements into a standard
proportional-integral-derivative (PID) control loop with an overall
gain less than unity to provide damping and avoid overshoot and
oscillation during guiding.

To implement the defocus observing mode, the image analysis part of
this astrometric routine was replaced with a custom source detection
routine using a standard matched filter approach
(e.g. \citealt{1985MNRAS.214..575I}), where in the case of defocused
images, rather than using a standard approximately Gaussian filter
kernel, the filter kernel was instead a model of the defocused
telescope PSF.  This technique is appropriate for analysis of images
with mild amounts of defocus, such as needed on MEarth.  Previous work
(e.g. \citealt{2013PASP..125..548M}) has usually concentrated on the
case of severe defocus, where different analysis techniques are
needed.

The PSF model was constructed by approximating the telescope entrance
pupil as a circular annulus, and introducing defocus by setting the
complex phase of this function to a multiple of the $Z_2^0$ Zernike
mode.  The resulting PSF was computed by taking the inverse Fourier
transform of this function.  In practice, it was also convolved by a
Moffat profile \citep{1969A&A.....3..455M} with parameters chosen
based on standard in-focus MEarth observations to approximate seeing
and any effects other than diffraction that contribute to the system's
normal in-focus PSF spot size.  The relationship between the $Z_2^0$
Zernike coefficient and HFD was determined empirically.

The PSF model was also used to compute exposure times and set
photometric aperture radii for the automatic extraction pipeline.  We
found that these theoretical estimates of exposure times
based on the idealised PSF models were rather optimistic, and in
practice it was necessary to use shorter exposures (or equivalently,
somewhat more defocus for a given desired exposure time) to avoid the
risk of saturation due to non-uniformity of the resulting defocused
star image.  This can be caused by atmospheric turbulence
(particularly in short exposures), but also other optical aberrations
affecting the defocused star image, such as coma, which causes an
asymmetric distribution of brightness around the resulting ring shaped
PSF, and can cause one side of the ring to become too bright.  Being
remotely operated robotic telescopes, it was not always possible to
maintain optimal collimation of the MEarth telescope optics, and while
this had minimal effect on the normal in-focus images used for the
majority of the survey, it did noticeably affect the defocused PSFs.

With an appropriate detection threshold, this source detection
procedure was found to produce quite robust results, albeit at 
reduced sensitivity to faint sources, and with a practical upper limit
to the defocus HFD of approximately 15 pixels.  Given the field of
view of the MEarth telescopes of approximately 27x27 arcmin the number
of detected sources was found to still be sufficient for accurate
multi-star guiding using the astrometric solutions on nearly all of
the targets observed over several years of observations, including
hundreds of TESS objects of interest.


\begin{table}
\centering
\footnotesize
\caption{Literature values for K2-2.}
\begin{tabular}{lcc}
\hline
&Other Identifiers &\\ \hline
&TIC 422618449 &\\
&2MASS J23354927+0026436 & \\
&EPIC 60021410 &\\
&WISE J233549.11+002641.9 & \\
\hline
Parameter &  Description & Value \\
\hline
$\alpha_\mathrm{J2000}$ & Right ascension (R.A.) & 23:35:49.29   \\
$\delta_\mathrm{J2000}$ & Declination (Dec.) & 00:26:43.84   \\
\\
$G$ & Gaia EDR3 $G$ mag &	9.932 $\pm$ 0.020	\\
$G_\mathrm{Bp}$ & Gaia EDR3 $B_P$ mag &	10.393 $\pm$ 0.020	 \\
$G_\mathrm{Rp}$ & Gaia EDR3 $R_P$ mag & 9.317 $\pm$ 0.020	\\ 
$T$ & \textit{TESS} mag &  	9.374 $\pm$ 0.006 \\
$J$ & 2MASS $J$ mag &	8.604 $\pm$ 0.021  \\
$H$ & 2MASS $H$ mag &	 8.140 $\pm$ 0.033 \\
$K_S$ & 2MASS $K_S$ mag &	8.029 $\pm$ 0.021\\
WISE1 & WISE1 mag & 	7.996 $\pm$ 0.030  \\
WISE2 & WISE2 mag & 	8.078 $\pm$ 0.030 	 \\
WISE3 & WISE3 mag &	 8.019 $\pm$ 0.030 \\
WISE4 & WISE4 mag & 7.878 $\pm$ 0.199  \\
\\
$\mu_\alpha$ &Gaia p.m. in R.A. & -232.90 $\pm$ 	0.019\\ 
$\mu_\delta$ & Gaia p.m. in Dec.  & -187.$\pm$ 0.017 \\
$\pi$ & Gaia parallax (mas) & $16.004\pm0.046$  \\
\hline
\end{tabular}
 \begin{flushleft} 
  \footnotesize{
    \textbf{Notes.} The uncertainties of the photometry have a systematic error floor applied. Proper motions taken from the Gaia EDR3 archive and are in J2016. Parallaxes from Gaia EDR3 have a correction applied according to \cite{Lindegren:2021}. 
    }
\end{flushleft}
\label{tab:literature}
\end{table}

\subsection{ULMT}

Once the ephemeris was refined from the MEarth observations, an ingress of K2-2 b was observed using the University of Louisville Manner Telescope (ULMT; formerly MVRC) at the Mt. Lemmon summit of Steward Observatory, Arizona. The observation was made in the $r^\prime$ band with 50 second exposure time on UT 2016 October 10. The setup used for the observation included a 0.6 m f/8 RC Optical Systems Ritchey–Chr\'etien telescope and SBIG STX-16803 CCD camera with a 4k$\times$4k array of 9 $\mu$m pixels, which yielded a 26.6' $\times$ 26.6' field of view and 0.39 pixel-1 plate scale. The images were calibrated and photometric data were extracted using {\tt AstroImageJ}~ \citep{Collins:2017}, and the light curves were detrended against airmass in the global fit.

\subsection{Spitzer}
\label{sec:obsSpitzer}

With the ephemeris more precisely constrained from the MEarth and ULMT transits, Spitzer was used to observe a single transit of K2-2 b on UT 2017 April 1 (P.I. M. Werner, observing program 13052, AOR 62428416; \citealp{Werner:2016-spitzer}). The observation was 10.5 hours long, and was taken with the InfraRed Array Camera (IRAC; \citealt{Fazio:2004}) channel 2 (4.5 µm) with a 2-second exposure time. We used the technique described in \citet{Livingston:2018} to extract the light curve. In brief, we extracted an optimal light curve by selecting the photometric aperture that minimized both white and red noise, and then corrected for systematics using pixel-level decorrelation (PLD; \citealt{Deming:2015}). 

As Spitzer can have correlated noise due to spacecraft systematics, we scaled the per point errors so that we did not underestimate the uncertainties. We followed the procedure from \citet{Winn:2008}, where a scaling factor, $\beta$, is applied to the measured standard deviation to account for time-correlated noise. We first calculated the out-of-transit standard deviation for the unbinned data, $\sigma_1$ (for this calculation we conservatively defined out-of-transit as being outside of a full transit duration centered at the transit midpoint). We then binned the out-of-transit data points to a series of 10 temporal bin widths ranging from 4.2 minutes to 8.8 minutes, increasing in equal steps of 0.46 minutes. The limits on the bin widths correspond to the 1$\sigma$ range of the ingress/egress duration based on a preliminary fit using K2 and TESS light curves. 

We then calculated the standard deviation for each set of binned data. In general, this should be equivalent to $\sigma_N = \sigma_1/\sqrt{N}\times \sqrt{M/(M-1)}$, where M is number of bins and N is data points per bin, if there is no time-correlated noise. However, the measured $\sigma_N$ can be larger than the expected value (by the factor $\beta$). We calculated this factor for each bin width, then used the mean value across all widths as the final value for $\beta$. Finally, we scaled the original unbinned, out-of-transit error bars by the factor $\beta=1.19$, which is used as the per point uncertainty in our global fit.


\subsection{{\it TESS} Photometry}
\label{sec:TESS}
A single transit was observed by the Transiting Exoplanet Survey Satellite (TESS) in each of Sectors 42 and 70. We used the 120 second cadence lightcurves in our global fits. We retrieved the light curve through the Python package \textit{Lightkurve} \citep{Lightkurve:2018}, selecting the light curve processed through the Science Processing Operations Center (SPOC) pipeline at the NASA Ames Research Center \citep{Jenkins:2016}, which corrects for various systematics and identifies transits. The light curves were created from the Pre-search Data Conditioned Simple Aperture Photometry (PDCSAP) flux, which uses the optimal TESS aperture to extract the flux and corrects the target for systematics using the PDC module \citep{Stumpe:2012,Stumpe:2014,Smith:2012}. To correct for stellar variability and any remaining systematics based on the out-of-transit photometry, we used the spline-fitting routine {{\tt keplerspline}}\footnote{\url{https://github.com/avanderburg/keplerspline}}  \citep{Vanderburg:2014}. We applied an initial estimate on the per-point errors for the corrected light curves as being the median absolute deviation of the out-of-transit photometry. We note that the per-point error is optimized through a fitted jitter term in the \texttt{EXOFASTv2} global fit (See Section \ref{sec:GlobalFit}).

\begin{table}[t]
\centering
\footnotesize
\caption{Photometry used in this analysis.}
\begin{tabular}{lllc}

\hline

Observatory & Date & Filter & Cadence  \\ 
\hline
K2 & February 6 2014 &  \textit{Kepler} &  30 min  \\
MEarth South& September 21 2016  & $i'$  &1 min    \\
MEarth South, North & September 30 2016  & $i'$  &1 min  \\
MEarth North & October 9 2016  & $i'$  &  1 min  \\
ULMT & October 10 2016  & $r'$  & 50 sec  \\
Spitzer & April 1 2017   &  4.5$\mu$m & 2 sec    \\
TESS & August 21 2021 & \textit{TESS}  &  2 min \\
TESS & September 21 2023 & \textit{TESS}  & 2 min \\
\hline
\end{tabular}

\begin{flushleft}
 \footnotesize{ \textbf{\textsc{Notes:}} Each telescope caught one full transit, except for ULMT which observed the ingress and partial transit. Observations with MEarth North used Telescopes 1, 2, 3 and 6, while MEarth South included Telescopes 11, 12, 16 and 18.  
}
\end{flushleft}
\label{tab:obs}
\end{table}

\begin{figure*}
    \centering
    \includegraphics[width=\linewidth]{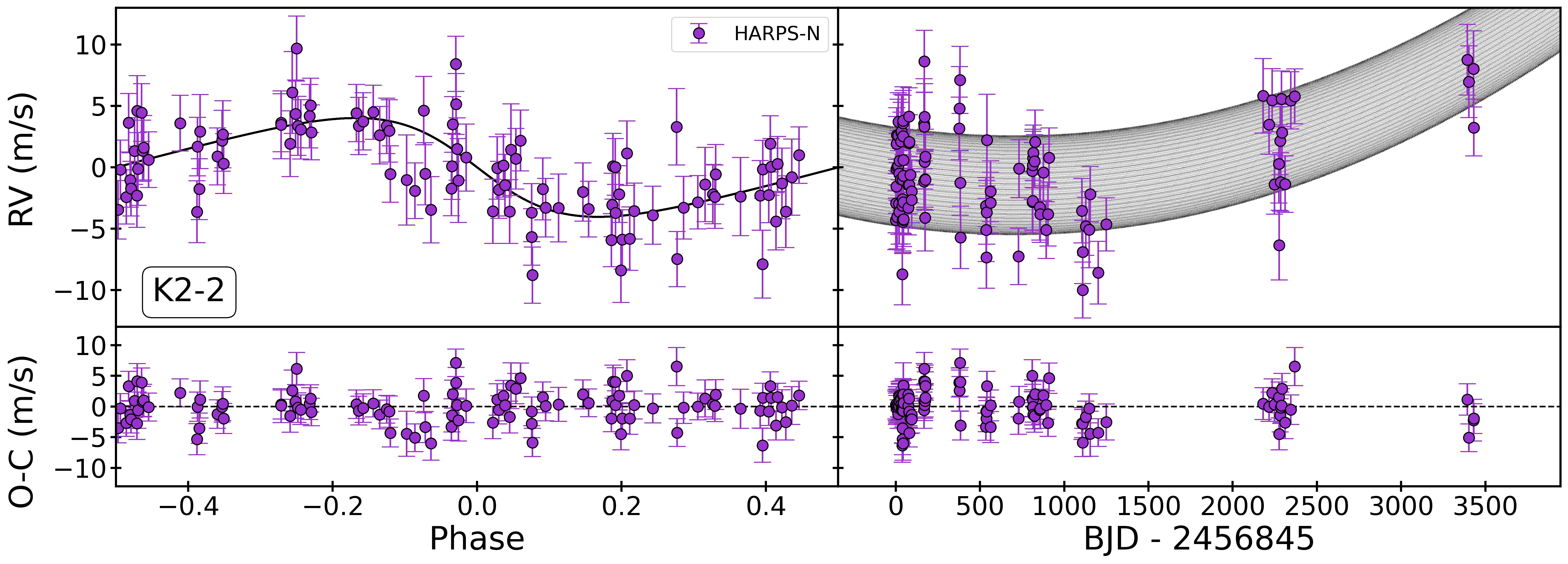}
    \caption{Archival HARPS-N radial velocities for K2-2 from  \citet{Vanderburg:2015-K2-2} and \citet{Bonomo:2023}. The left panel shows the phased-folded RVs, and the right panel shows the long-term trend in the unphased RVs. }
    \label{fig:rvs}
\end{figure*}


\subsection{Archival Spectroscopy}
\label{sec:RVs}

We included archival spectroscopy to determine the host star properties and to refine the mass measurement of K2-2 b. In particular, to better characterize the host star in the global fit, we used metallicity measurements of K2-2 from the Tillinghast Reflector Echelle Spectrograph (TRES; \citealt{furesz:2008}) on the 1.5m Tillinghast Reflector at the Fred L. Whipple Observatory (FLWO). This is in keeping with our procedure for the larger Synergy catalog, where we are using TRES metallicities where available. The stellar parameters using TRES spectra were derived using the Stellar Parameter Classification (SPC; \citealt{Buchhave:2012}). Three measurements from TRES ([M/H] = -0.193 $\pm$ 0.086, -0.191 $\pm$ 0.08, 0.009 $\pm$ 0.08) were available through the ExoFOP website\footnote{\url{https://exofop.ipac.caltech.edu/tess/target.php?id=422618449}}. We used the mean value to place a Gaussian prior on metallicity ([Fe/H]) of -0.125 $\pm$ 0.08.

We used a total of 105 spectra of K2-2, including those used in \citet{Vanderburg:2015-K2-2} and \citet{Bonomo:2023}, acquired using the High Accuracy Radial velocity Planet Searcher for the Northern hemisphere (HARPS-N) on the 3.6m Telescopio Nazionale Galileo (TNG) at the Roque de los Muchachos Observatory \citep{Consentino:2012}, in order to better characterize the mass of K2-2 b (Figure \ref{fig:rvs}). Each observation had either 15 or 30 minutes exposure time, with a resolving power of $R=$ 115,000. We followed the procedure of \citet{Dumusque:2021} to reduce the RVs that were used in our global fits. The observations occurred in two main blocks, separated by $\sim2.5$ years; the first run was from  UT 2014 July 7 to December 6 2017, and the second from UT 2020 June 25 to 2023 November 27. The second series of RVs was significantly offset to the earlier measurements, which led us to apply post-processing systematics corrections to investigate whether the offset was instrumental or physical in nature. 


\subsubsection{YARARA processing to correct remaining systematics}

YARARA \citep{yarara1:2021} is a post-processing methodology that aims to perform correction of the spectra by the analysis of the spectra time-series. While a more advanced version of the pipeline has been presented recently in \cite{yarara2:2023} (sometimes referred to as the YARARA V2 or YV2 datasets), the SNR of the target was too low to apply those advanced methods of correction (such as the SHELL presented in \cite{cretignier:2022}) and we remained with the YARARA V1 or YV1 version of the products. 

The corrections available in YARARA cover as much as the telluric lines, as instrumental systematics or stellar activity. The pipeline usually starts from the S1D order-merged spectra produced by official DRS that have been continuum normalized by RASSINE \citep{rassine:2020}. The method then consists of a multi-linear decorrelation by fitting a basis of vectors that are designed to correct for some dedicated effects, either obtained by optimized extraction (see e.g. \cite{stalport:2023}) or by principal component analysis (PCA) as initially presented in \cite{yarara1:2021}. For a dataset around SNR $\sim$50, the main corrections that are possible to perform consist of removing cosmic, telluric lines, and the change of the instrumental PSF \citep{stalport:2023}. Even if a clear and strong emission is detected in the core of the CaII H\&K lines, no reliable and precise extraction of the signal could be achieved and the stellar activity correction that mainly relies on this proxy (which contains most of the information from active regions \citep{cretignier:2024}) was therefore skipped. The RVs were obtained with a cross-correlation function (CCF) on the corrected spectra using a line list optimised for the star following the line centre procedure described in \cite{cretignier:2020a}.

After the application of YARARA, we still detect the long-trend signal which discards any potential effects from telluric or change of the instrumental PSF at the precision level of our data.

\begin{figure}
    \centering
    \includegraphics[width=\linewidth]{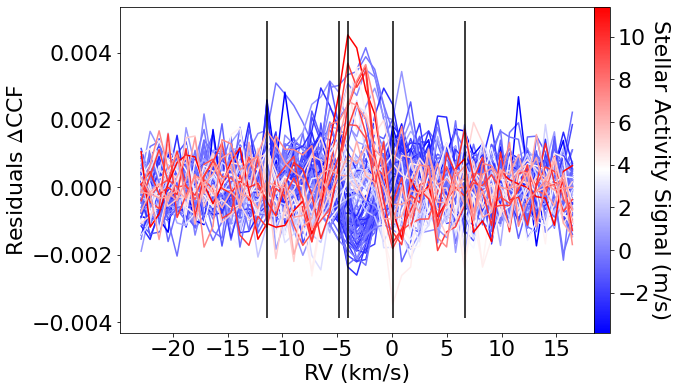}
    \caption{Residual CCFs ($\Delta$CCFs) computed from HARPS-N spectra. The residual CCFs are computed by subtracting a median CCF. The CALM model-predicted stellar activity signal is indicated by the color (red = redshifted RVs, blue = blue-shifted RVs). The 5 CCF indexes used in our stellar activity model are indicated by black lines. }
    \label{fig:ccf_indexes}
\end{figure}

\begin{figure*}
    \centering
    \includegraphics[width=0.9\linewidth]{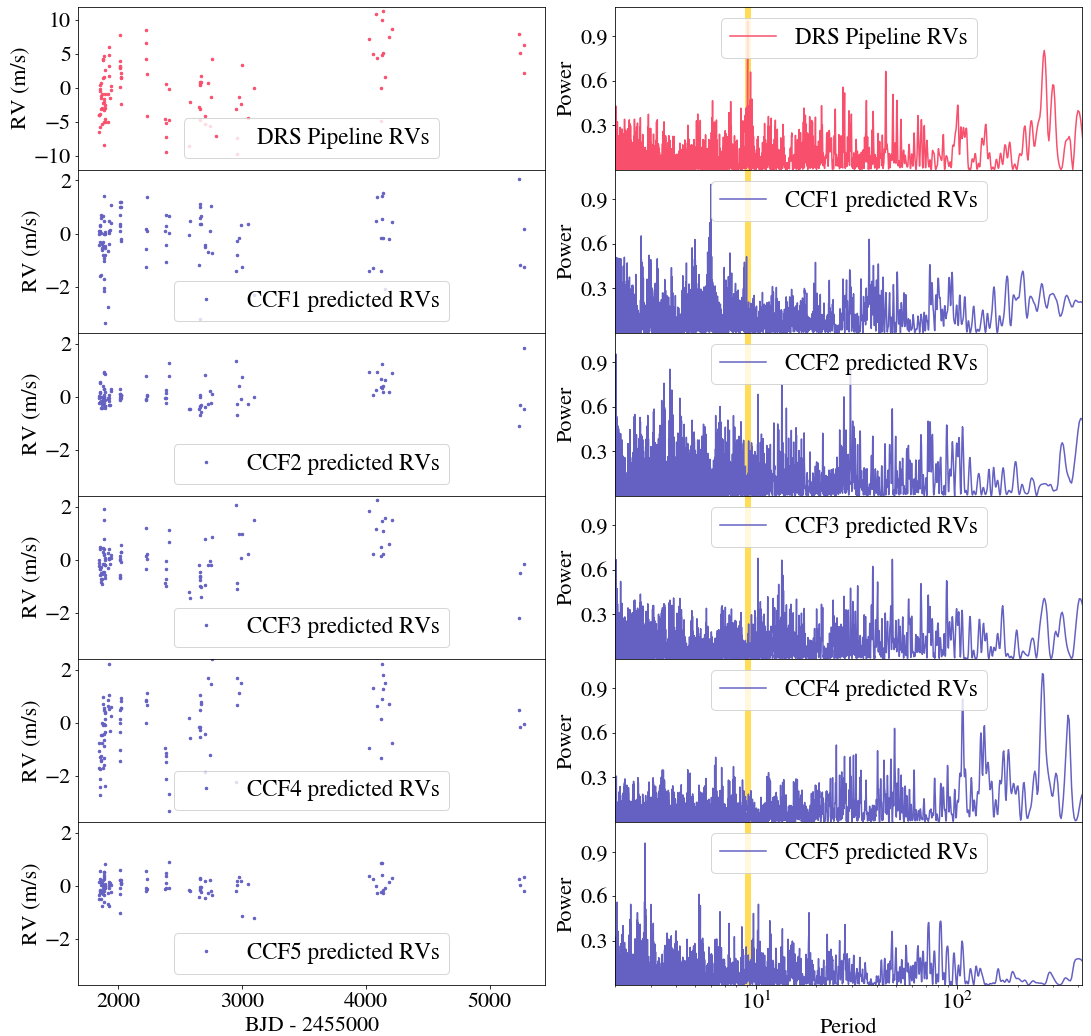}
    \caption{Timeseries and periodograms of the CALM predicted stellar variability. In the left panels, the DRS pipeline radial velocities and the stellar variability predictions from CCF index 1, 2, 3, 4, and 5 are plotted as a function of time. The location of these CCF indexes are indicated in Figure \ref{fig:ccf_indexes}. On the right panel, the Lomb-Scargle periodograms of the corresponding RV timeseries are plotted. In yellow, the Keplerian period of K2-2 is indicated in the periodograms. We do not see signals at this planetary period, which provides reassurance that CALM is not absorbing or creating planetary signals.}
    \label{fig:ccf_pgram}
\end{figure*}


\subsubsection{CCF Activity Linear Model (CALM) to model stellar variability}
\label{sec:ccf}

    To model stellar variability in the radial velocities, we used activity indicators derived using the CCF Activity Linear Model (CALM)  \citep{deBeurs2024}. CALM is a linear regression method which exploits the shape changes that stellar variability introduces into the cross-correlation functions (CCFs) computed from stellar spectra. Since CCFs represent an average of all line shapes in a star's spectrum, CALM is especially sensitive to line shape changes that persist in most spectral lines. In this method, we do not include the entire CCF in our model since CCFs are comprised of 49-element arrays and we only have 105 RVs. Including the entire CCF would lead to overfitting. We experimented with sampling various fractions of the CCFs and across random locations within the CCF.  We found that using 5 CCF locations provides a balance between preventing overfitting and optimizing goodness-of-fit. These 5 CCF locations are then used to decorrelate against in the global fit performed using \texttt{EXOFASTv2}. We visualize the CCFs for K2-2 and the specific 5 CCF locations in Figure \ref{fig:ccf_indexes}, where we observe a clear pattern in the stellar variability and the CCF shape changes. This pattern allows us to use CALM to probe and predict stellar activity contributions to the RVs. In Figure \ref{fig:ccf_pgram}, we plot the CALM model predicted stellar activity contributions to the RVs both in time and in the fourier domain. These activity indexes are able to probe both short- and long-term activity signals while preserving the planetary reflex motion. The $\sim$270 day signal that is predicted by the CCF4 parameter was also found by \citet{Bonomo:2023} and they noted that this signal is also seen in the periodograms of s-index and FWHM. This suggests that this signal corresponds to stellar variability and may be on a timescale longer than the stellar rotation period for K2-2. 

\section{Global Fits} 
\label{sec:GlobalFit}

Following the method described in \citet{Thygesen:2023}, we used the differential evolution Markov Chain Monte Carlo (DE-MCMC) exoplanet fitting software \texttt{EXOFASTv2} \citep{Eastman:2013,Eastman:2019} to simultaneously fit the parameters of K2-2 b and its host star. For a global fit to be accepted as converged, we required that the Gelmin-Rubin statistic be less that 1.01 and the number of independent draws, $T_z$, greater than 1000. The global fits use MCMC sampling to find the best fit parameters for the system based on the photometric and spectroscopic data. We placed priors on several parameters as follows: a uniform prior from 0 to an upper bound of 0.09858 on the line-of-sight extinction ($A_v$) from \citet{Schlegel:1998} and \citet{Schlafly:2011}; a Gaussian prior on parallax of 16.0044 $\pm$ 0.0456 from Gaia Early Data Release 3 (accounting for the small systematic offset; EDR3; \citealt{Gaia:2016,GaiaEDR3:2021,Lindegren:2021}); and a Gaussian prior on metallicity ([Fe/H]) of -0.125 $\pm$ 0.08 based on measurements from TRES (see Section \ref{sec:RVs}). The fit also included the spectral energy distribution (SED) photometry as reported by Gaia EDR3 \citep{GaiaEDR3:2021}, WISE \citep{Cutri:2012} and 2MASS \citep{Cutri:2003} (see Table \ref{tab:literature}). To better characerize the host star, the MESA Isochrones and Stellar Tracks (MIST) stellar evolution models \citep{Paxton:2011, Paxton:2013, Paxton:2015, Choi:2016, Dotter:2016} were used within the \texttt{EXOFASTv2} fits. Within {\tt EXOFASTv2}, limb darkening is constrained via priors derived from models by \cite{Claret:2011} and \cite{Claret:2017}, with physical bounds from \cite{Kipping:2013} (see Section 3 of \cite{Eastman:2019} for more details on how {\tt EXOFASTv2} constrains limb darkening).

Although the TESS PDCSAP light curves generally have a correction applied for any contaminating sources, we fitted for a dilution term in case of any sources that may have been missed, based on the contamination ratio (CR) for K2-2 of 0.002101 as reported in the TESS input catalog (TICv8, \citealp{Stassun:2018_TIC}). We used placed a 10\% Gaussian prior on the dilution centered about CR/(1+CR) = 0.0021. However, the fitted dilution was consistent with zero in all the fits we ran. 

To account for any residual correlated noise in the systematics-corrected Spitzer data within the \texttt{EXOFASTv2} fit (see Section \ref{sec:obsSpitzer}), we followed the procedure outlined in \S3 of \citet{Rodriguez:2020}. We scaled the uncertainties by the factor $\beta=1.19$ before using the light curve in the global fit. To ensure \texttt{EXOFASTv2} did not reduce the per-point uncertainties on the Spitzer photometry within the fit, we enforced a lower bound on the variance of zero, otherwise the global fit could over-correct the scaled uncertainties to be consistent with pure white noise. 

\subsection{RV model selection}

As the RVs still exhibited an offset in the second observing block after all processing (see Section \ref{sec:RVs}), we compared five different models that attempt to model this long-term change and evaluated their goodness-of-fit with \texttt{EXOFASTv2}, while keeping all other inputs and priors the same. For each of these models, we first performed a fit using CALM since these long-term trends could be caused by stellar variability. We then took the initial CALM fit to the RVs for each model and ran a global fit with \texttt{EXOFASTv2}. The five models are listed in Table \ref{tab:models} and each include the CALM model, but differ in their modeling of the long-term trends where they include some combination of a linear  ($\dot{\gamma}$) trend with time, a quadratic ($\ddot{\gamma}$) trend with time, and/or an offset $D$ between the two observing blocks. In particular, our models include (i) a CALM model with a linear and quadratic trend with time that treats the RV timeseries as one RV observing season without an offsets between the two observing blocks, (ii) a CALM and linear trend model that treats the RV timeseries as one RV observing season without an offset, (iii) a CALM model with an offset $D$ between the two observing blocks, (iv) a CALM model with a linear trend and an offset $D$, and (v) a CALM model with a linear and quadratic trend and an offset $D$ . For the models where we treated the two observing blocks as separate seasons, this allows for different zero-points to be determined for each season. Comparing the Bayesian Information Criterion (BIC) of the models, we found that those including an offset component (i.e. two observing seasons) are heavily disfavored as seen in Table \ref{tab:models}. The single-season models perform comparabley and we adopt the quadratic-trend model as it has the lowest BIC. 


\begin{table}[t]
\centering
\footnotesize
\caption{Models tested for long-term RV trend.}
\begin{tabular}{lll}

\hline

Model & Description & $\Delta$BIC  \\ 
\hline
(i) & One RV season, linear and quadratic trend with time & 0.0  \\
(ii) & One RV season, linear trend with time &  0.72 \\
(iii) & Two RV seasons, no long-term trend & 49.75 \\
(iv) & Two RV seasons, linear trend with time& 55.15 \\
(v)  & Two RV seasons, linear and quadratic trend with time & 68.21 \\
\hline
\end{tabular}
\label{tab:models}
\end{table}



\section{Results and Discussion}

\label{sec:discussion}

\begin{figure}
    \centering
    \includegraphics[width=\linewidth]{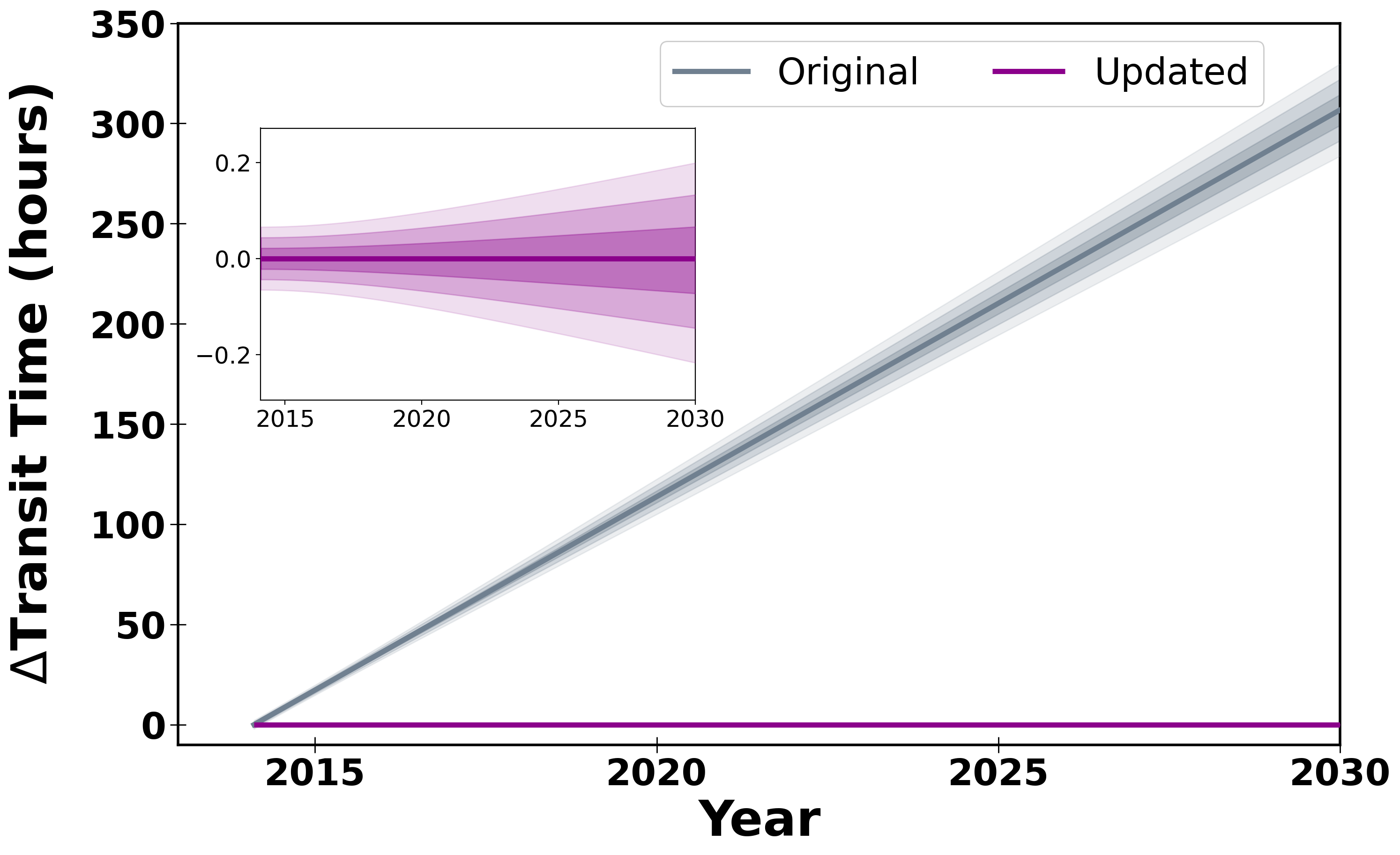}
    \caption{Projected difference in the time of transit for K2-2 b to the year 2030 using the original ephemeris (gray) and the new ephemeris from this work (purple). Shaded regions indicating up to the 3$\sigma$ level uncertainty are shown. The inset shows the updated ephemeris, zoomed in for clarity.}
    \label{fig:ephemeris}
\end{figure}

\providecommand{\bjdtdb}{\ensuremath{\rm {BJD_{TDB}}}}
\providecommand{\feh}{\ensuremath{\left[{\rm Fe}/{\rm H}\right]}}
\providecommand{\teff}{\ensuremath{T_{\rm eff}}}
\providecommand{\teq}{\ensuremath{T_{\rm eq}}}
\providecommand{\ecosw}{\ensuremath{e\cos{\omega_*}}}
\providecommand{\esinw}{\ensuremath{e\sin{\omega_*}}}
\providecommand{\msun}{\ensuremath{\,M_\Sun}}
\providecommand{\rsun}{\ensuremath{\,R_\Sun}}
\providecommand{\lsun}{\ensuremath{\,L_\Sun}}
\providecommand{\mj}{\ensuremath{\,M_{\rm J}}}
\providecommand{\rj}{\ensuremath{\,R_{\rm J}}}
\providecommand{\me}{\ensuremath{\,M_{\rm E}}}
\providecommand{\re}{\ensuremath{\,R_{\rm E}}}
\providecommand{\fave}{\langle F \rangle}
\providecommand{\fluxcgs}{10$^9$ erg s$^{-1}$ cm$^{-2}$}

\begin{deluxetable}{llc}
\tablecaption{Median values and 68$\%$ confidence interval for K2-2 stellar parameters from the EXOFASTv2 global fit.}
\tabletypesize{\footnotesize}
\tablehead{\colhead{~~~Parameter} & \colhead{Units} & \colhead{Values}}
\startdata
Priors: & & \\
~~~~$\pi$\dotfill &Gaia parallax (mas)\dotfill &  $\mathcal{G}[16.0044, 0.0465]$  \\
~~~~$[{\rm Fe/H}]$\dotfill & Metallicity (dex)\dotfill & $\mathcal{G}[-0.125, 0.080]$ \\
~~~~$A_V$\dotfill & $V$-band extinction (mag)\dotfill & $\mathcal{U}[0, 0.0985]$  \\ \hline
~~~~$M_*$\dotfill &Mass (\msun)\dotfill &$0.800^{+0.033}_{-0.030}$\\
~~~~$R_*$\dotfill &Radius (\rsun)\dotfill &$0.758^{+0.024}_{-0.022}$\\
~~~~$L_*$\dotfill &Luminosity (\lsun)\dotfill &$0.3364^{+0.0100}_{-0.0096}$\\
~~~~$F_{Bol}$\dotfill &Bolometric Flux $\times 10^{-9}$ (cgs)\dotfill &$2.757^{+0.081}_{-0.076}$\\
~~~~$\rho_*$\dotfill &Density (cgs)\dotfill &$2.59^{+0.26}_{-0.24}$\\
~~~~$\log{g}$\dotfill &Surface gravity (cgs)\dotfill &$4.582^{+0.030}_{-0.031}$\\
~~~~$T_{\rm eff}$\dotfill &Effective Temperature (K)\dotfill &$5048^{+79}_{-78}$\\
~~~~$[{\rm Fe/H}]$\dotfill &Metallicity (dex)\dotfill &$0.000^{+0.045}_{-0.039}$\\
~~~~$[{\rm Fe/H}]_{0}$\dotfill &Initial Metallicity$^{1}$ \dotfill &$0.000\pm0.055$\\
~~~~$Age$\dotfill &Age (Gyr)\dotfill &$5.5^{+5.0}_{-3.9}$\\
~~~~$EEP$\dotfill &Equal Evolutionary Phase$^{2}$ \dotfill &$335^{+16}_{-34}$\\
~~~~$A_V$\dotfill &V-band extinction (mag)\dotfill &$0.045^{+0.035}_{-0.031}$\\
~~~~$\sigma_{SED}$\dotfill &SED photometry error scaling \dotfill &$0.76^{+0.32}_{-0.19}$\\
~~~~$\varpi$\dotfill &Parallax (mas)\dotfill &$16.004\pm0.046$\\
~~~~$d$\dotfill &Distance (pc)\dotfill &$62.48\pm0.18$\\
\enddata
\label{tab:exo_star}
\begin{flushleft} 
  \footnotesize{
    \textbf{Notes.} See Table 3 in \citet{Eastman:2019} for a detailed description of all parameters. Gaussian and uniform priors are indicated as $\mathcal{G\mathrm{[mean, \sigma]}}$ and $\mathcal{U\mathrm{[lower~bound, upper~bound]}}$, respectively. The metallicity prior is adopted from the average of three TRES measurements: [M/H] = -0.193, -0.191, 0.009 (see Section \ref{sec:RVs} for details). 
    $^1$The metallicity of the star at birth. $^2$Corresponds to static points in a star's evolutionary history. See \S2 in \citet{Dotter:2016}.
    }
\end{flushleft}
\end{deluxetable}

\providecommand{\bjdtdb}{\ensuremath{\rm {BJD_{TDB}}}}
\providecommand{\feh}{\ensuremath{\left[{\rm Fe}/{\rm H}\right]}}
\providecommand{\teff}{\ensuremath{T_{\rm eff}}}
\providecommand{\teq}{\ensuremath{T_{\rm eq}}}
\providecommand{\ecosw}{\ensuremath{e\cos{\omega_*}}}
\providecommand{\esinw}{\ensuremath{e\sin{\omega_*}}}
\providecommand{\msun}{\ensuremath{\,M_\Sun}}
\providecommand{\rsun}{\ensuremath{\,R_\Sun}}
\providecommand{\lsun}{\ensuremath{\,L_\Sun}}
\providecommand{\mj}{\ensuremath{\,M_{\rm J}}}
\providecommand{\rj}{\ensuremath{\,R_{\rm J}}}
\providecommand{\me}{\ensuremath{\,M_{\rm E}}}
\providecommand{\re}{\ensuremath{\,R_{\rm E}}}
\providecommand{\fave}{\langle F \rangle}
\providecommand{\fluxcgs}{10$^9$ erg s$^{-1}$ cm$^{-2}$}

\begin{deluxetable*}{lcc}
\tablecaption{Median values and 68$\%$ confidence interval for K2-2 b planetary parameters from the EXOFASTv2 global fit.}
\tabletypesize{\footnotesize}
\tablehead{\colhead{~~~Parameter} & \colhead{Units} & \colhead{Values}}
\startdata
~~~~$P$\dotfill &Period (days)\dotfill &$9.1004157^{+0.0000041}_{-0.0000045}$\\
~~~~$R_P$\dotfill &Radius (\re)\dotfill &$2.469^{+0.10}_{-0.091}$\\
~~~~$M_P$\dotfill &Mass (\me)\dotfill &$9.7\pm1.2$\\
~~~~$T_0$\dotfill &Optimal conjunction Time$^{1}$ (\bjdtdb)\dotfill &$2458072.29291^{+0.00062}_{-0.00061}$\\
~~~~$a$\dotfill &Semi-major axis (AU)\dotfill &$0.0792^{+0.0011}_{-0.0010}$\\
~~~~$i$\dotfill &Inclination (Degrees)\dotfill &$88.91^{+0.68}_{-0.45}$\\
~~~~$e$\dotfill &Eccentricity$^2$ \dotfill &$0.215^{+0.056}_{-0.094}$\\
~~~~$\omega_*$\dotfill &Argument of Periastron (Degrees)\dotfill &$88^{+19}_{-20}$\\
~~~~$T_{eq}$\dotfill &Equilibrium temperature$^{3}$ (K)\dotfill &$753.2^{+7.1}_{-6.9}$\\
~~~~$\tau_{\rm circ}$\dotfill &Tidal circularization timescale (Gyr)\dotfill &$1310^{+540}_{-430}$\\
~~~~$K$\dotfill &RV semi-amplitude (m/s)\dotfill &$3.54\pm0.42$\\
~~~~$R_P/R_*$\dotfill &Radius of planet in stellar radii \dotfill &$0.02981^{+0.00079}_{-0.00061}$\\
~~~~$a/R_*$\dotfill &Semi-major axis in stellar radii \dotfill &$22.46\pm0.72$\\
~~~~$\delta$\dotfill &$\left(R_P/R_*\right)^2$ \dotfill &$0.000889^{+0.000048}_{-0.000036}$\\
~~~~$\delta_{\rm Kepler}$\dotfill &Transit depth in Kepler (fraction)\dotfill &$0.001186^{+0.000052}_{-0.000050}$\\
~~~~$\delta_{\rm i'}$\dotfill &Transit depth in i' (fraction)\dotfill &$0.001092^{+0.000035}_{-0.000034}$\\
~~~~$\delta_{\rm r'}$\dotfill &Transit depth in r' (fraction)\dotfill &$0.001171^{+0.000056}_{-0.000051}$\\
~~~~$\delta_{\rm 4.5\mu m}$\dotfill &Transit depth in $4.5\mu m$ (fraction)\dotfill &$0.000922^{+0.000046}_{-0.000040}$\\
~~~~$\delta_{\rm TESS}$\dotfill &Transit depth in TESS (fraction)\dotfill &$0.001092^{+0.000039}_{-0.000038}$\\
~~~~$\tau$\dotfill &Ingress/egress transit duration (days)\dotfill &$0.00329^{+0.00088}_{-0.00036}$\\
~~~~$T_{14}$\dotfill &Total transit duration (days)\dotfill &$0.1013^{+0.0015}_{-0.0014}$\\
~~~~$T_{FWHM}$\dotfill &FWHM transit duration (days)\dotfill &$0.0978\pm0.0013$\\
~~~~$b$\dotfill &Transit Impact parameter \dotfill &$0.34^{+0.20}_{-0.22}$\\
~~~~$b_S$\dotfill &Eclipse impact parameter \dotfill &$0.51^{+0.15}_{-0.31}$\\
~~~~$\tau_S$\dotfill &Ingress/egress eclipse duration (days)\dotfill &$0.00540^{+0.00074}_{-0.00051}$\\
~~~~$T_{S,14}$\dotfill &Total eclipse duration (days)\dotfill &$0.141^{+0.027}_{-0.028}$\\
~~~~$T_{S,FWHM}$\dotfill &FWHM eclipse duration (days)\dotfill &$0.135^{+0.027}_{-0.028}$\\
~~~~$\delta_{S,2.5\mu m}$\dotfill &Blackbody eclipse depth at 2.5$\mu$m (ppm)\dotfill &$0.912^{+0.083}_{-0.073}$\\
~~~~$\delta_{S,5.0\mu m}$\dotfill &Blackbody eclipse depth at 5.0$\mu$m (ppm)\dotfill &$15.33^{+1.00}_{-0.85}$\\
~~~~$\delta_{S,7.5\mu m}$\dotfill &Blackbody eclipse depth at 7.5$\mu$m (ppm)\dotfill &$34.9^{+2.2}_{-1.7}$\\
~~~~$\rho_P$\dotfill &Density (cgs)\dotfill &$3.53^{+0.63}_{-0.57}$\\
~~~~$logg_P$\dotfill &Surface gravity \dotfill &$3.192^{+0.061}_{-0.065}$\\
~~~~$\Theta$\dotfill &Safronov Number \dotfill &$0.0274^{+0.0035}_{-0.0034}$\\
~~~~$\fave$\dotfill &Incident Flux (\fluxcgs)\dotfill &$0.0698^{+0.0032}_{-0.0029}$\\
~~~~$T_P$\dotfill &Time of Periastron (\bjdtdb)\dotfill &$2456689.01^{+0.30}_{-0.34}$\\
~~~~$T_S$\dotfill &Time of eclipse (\bjdtdb)\dotfill &$2456693.61^{+0.35}_{-0.36}$\\
~~~~$T_A$\dotfill &Time of Ascending Node (\bjdtdb)\dotfill &$2456705.54^{+0.20}_{-0.28}$\\
~~~~$T_D$\dotfill &Time of Descending Node (\bjdtdb)\dotfill &$2456690.73^{+0.32}_{-0.22}$\\
~~~~$V_c/V_e$\dotfill & \dotfill &$0.810^{+0.086}_{-0.047}$\\
~~~~$e\cos{\omega_*}$\dotfill & See footnote$^4$ \dotfill &$0.004^{+0.059}_{-0.060}$\\
~~~~$e\sin{\omega_*}$\dotfill & See footnote$^5$ \dotfill &$0.205^{+0.057}_{-0.098}$\\
~~~~$M_P/M_*$\dotfill &Mass ratio \dotfill &$0.0000365^{+0.0000044}_{-0.0000043}$\\
~~~~$d/R_*$\dotfill &Separation at mid transit \dotfill &$17.9^{+2.3}_{-1.7}$\\
\enddata
\label{tab:exo_planet}
\begin{flushleft} 
  \footnotesize{
    \textbf{Notes.} See Table 3 in \citet{Eastman:2019} for a detailed description of all parameters. \\
    %
$^1$Optimal time of conjunction minimizes the covariance between $T_C$ and Period. $^2$Note that due to the low significance of the eccentricity, this is consistent with $e=0$ when considering the Lucy-Sweeney bias \citep{Lucy:1971}. $^3$Assumes no albedo and perfect redistribution. $^{4, 5}$ Within the fits, these are parameterized as $\sqrt{e}\cos{\omega_*}$ and $\sqrt{e}\sin{\omega_*}$, respectively, to ensure a uniform prior on eccentricity.
    }
\end{flushleft}
\end{deluxetable*}

\begin{deluxetable}{llc}
\tablecaption{Median values and 68\% confidence interval for radial velocity parameters.}
\tabletypesize{\footnotesize}
\startdata
\multicolumn{2}{l}{Telescope Parameters:}&HARPS-N\smallskip\\
~~~~$\gamma_{\rm sys}$\dotfill & Systemic velocity (km/s) & $-2.91$ \\
~~~~$\gamma_{\rm rel}$\dotfill &Relative RV Offset (m/s)\dotfill &$0.02\pm0.63$\\
~~~~$\dot{\gamma}$\dotfill &RV slope (m/s/day)\dotfill &$0.00239\pm0.00039$\\
~~~~$\ddot{\gamma}$\dotfill &RV quadratic term (m/s/day$^2$)\dotfill &$0.00000133\pm0.00000036$\\
~~~~$\sigma_J$\dotfill &RV Jitter (m/s)\dotfill &$2.30^{+0.24}_{-0.22}$\\
~~~~$\sigma_J^2$\dotfill &RV Jitter Variance \dotfill &$5.27^{+1.2}_{-0.98}$\\
~~~~$CCF_{0}$\dotfill &Additive detrending coeff. \dotfill &$-3.23\pm0.92$\\
~~~~$CCF_{1}$\dotfill &Additive detrending coeff. \dotfill &$3.4\pm2.6$\\
~~~~$CCF_{2}$\dotfill &Additive detrending coeff. \dotfill &$2.0\pm2.0$\\
~~~~$CCF_{3}$\dotfill &Additive detrending coeff. \dotfill &$-6.6\pm1.5$\\
~~~~$CCF_{4}$\dotfill &Additive detrending coeff. \dotfill &$1.3\pm1.0$\\
\enddata
\label{tab:rv_params}
 \begin{flushleft} 
  \footnotesize{
    \textbf{Notes.} Reference epoch = 2458561.069744 BJD. Five additive detrending parameters were included to account for stellar activity (see Section \ref{sec:GlobalFit}). 
    }
\end{flushleft}
\end{deluxetable}

\begin{deluxetable*}{lccccc}
\centering
\tablecaption{Median values and 68\% confidence intervals for the photometric models.}
\tablehead{\omit}
\tabletypesize{\footnotesize}
\startdata
Telescope  &  \multicolumn{2}{c}{Wavelength Parameters} & \multicolumn{2}{c}{Transit Parameters} & Additive detrending coeff \\
  & $u_1^\dagger$ & $u_2^\ddagger$ & $\sigma^{2\star} (10^{-9})$ & $F_0^*$ & $C_0$ \\ \hline
 K2 & $0.57\pm0.052$ & $0.171\pm0.051$ & $2.45^{+0.56}_{-0.50}$ & $0.9999999\pm0.0000047$ & --- \\
MEarth (i') Telescope 1  & $0.426^{+0.022}_{-0.023}$ & $0.205\pm0.020$ & $2360^{+320}_{-290}$  &  $1.00042^{+0.000096}_{-0.000097}$ & $-0.00026\pm{0.00050}$ \\
MEarth (i') Telescope 2  &   $0.426^{+0.022}_{-0.023}$ & $0.205\pm0.020$    & $2220^{+290}_{-270}$  & $1.000525\pm0.000093$  & $0.00001\pm0.00049$  \\
MEarth (i') Telescope 3 &    $0.426^{+0.022}_{-0.023}$ & $0.205\pm0.020$    &  $4830^{+510}_{-470}$ & $1.00033\pm0.00012$  & $-0.00142\pm0.00057$ \\
 MEarth (i') Telescope 6  &   $0.426^{+0.022}_{-0.023}$ & $0.205\pm0.020$    & $4700^{+460}_{-430}$  &  $1.00053\pm0.00012$ & $-0.00076\pm0.00058$ \\
 MEarth (i') Telescope 11  &    $0.426^{+0.022}_{-0.023}$ & $0.205\pm0.020$    & $1310^{+200}_{-190}$  & $1.000213^{+0.000076}_{-0.000077}$  &  $-0.00147\pm0.00042$\\
 MEarth (i') Telescope 12  &   $0.426^{+0.022}_{-0.023}$ & $0.205\pm0.020$    & $2060^{+230}_{-210}$  & $1.000355\pm0.000080$  & $-0.00091\pm0.00045$ \\
 MEarth (i') Telescope 16  &    $0.426^{+0.022}_{-0.023}$ & $0.205\pm0.020$    & $940^{+160}_{-150}$  & $1.000344\pm0.000070$  &$-0.00003\pm0.00040$  \\
 MEarth (i') Telescope 18  &   $0.426^{+0.022}_{-0.023}$ & $0.205\pm0.020$  & $1530^{+200}_{-180}$ & $1.000253\pm0.000077$ & $-0.00098\pm0.00045$ \\
ULMT (r') & $0.551\pm0.054$ & $0.183\pm0.052$ & $3150.0^{+430.0}_{-380.0}$ & $0.99957\pm0.00014$ & $0.00041\pm0.00035$ \\
 Spitzer (4.5$\mu m$) & $0.077^{+0.047}_{-0.043}$ & $0.146\pm0.050$ & $5.7^{+9.2}_{-4.3}$   & $1.000003\pm0.000036$ & --- \\
 TESS Sector 42 &$0.428\pm0.038$ & $0.21\pm0.036$ & $28.3^{+8.1}_{-8.0}$ &  $1.0000131\pm0.0000073$ & --- \\
 TESS Sector 70 &$0.428\pm0.038$ & $0.21\pm0.036$ & $6.1\pm7.1$ &  $1.0000099\pm0.0000065$ & --- \\
 \enddata
\label{tab:wave}
 \begin{flushleft} 
  \footnotesize{
    \textbf{Notes.}$^\dagger$Linear limb-darkening coefficient. $^\ddagger$Quadratic limb-darkening coefficient. $^\star$Added variance. $^*$ Baseline flux. 
    }
\end{flushleft}
\end{deluxetable*}

In this work, we have combined multiple new observations with existing data available for K2-2 b to produce the most accurate and precise system parameters and transit ephemeris (transit time uncertainty $<$13 minutes in 2030). The period of K2-2 b has been updated to $9.1004157^{+4.1E-06}_{-4.5E-05}$ days and $T_0$ to $2458072.29291^{+0.00062}_{-0.00061}$ BJD (Figure \ref{fig:ephemeris}). The solutions for the stellar and planetary parameters are shown in Tables \ref{tab:exo_star} and \ref{tab:exo_planet}, respectively. Table \ref{tab:rv_params} contains the radial velocity parameters, including the detrending parameters we used, and Table \ref{tab:wave} lists the parameters of the photometric models for each light curve. We included the MOST light curve in a preliminary fit, as the transit window was observed four times in the full light curve. However, this did not add value to the fit, and the transit was not detectable even with the updated ephemeris, so we did not include the MOST data in the final global fit. The discovery period \citep{Vanderburg:2015-K2-2} we determined to be 28.8 minutes ($\sim$40 $\sigma$) from the true period. For context, if someone attempted an observation in 2025 of a K2-2 b transit using the original ephemeris, it would be $\sim$200 hours from the correct time. We note that this would only result in an offset of $\sim$18 hours from a transit of K2-2 b since the offset would be quite close to the orbital period of the planet by then, resulting in catching the next adjacent transit.

\begin{figure}
    \centering
    \includegraphics[width=\linewidth]{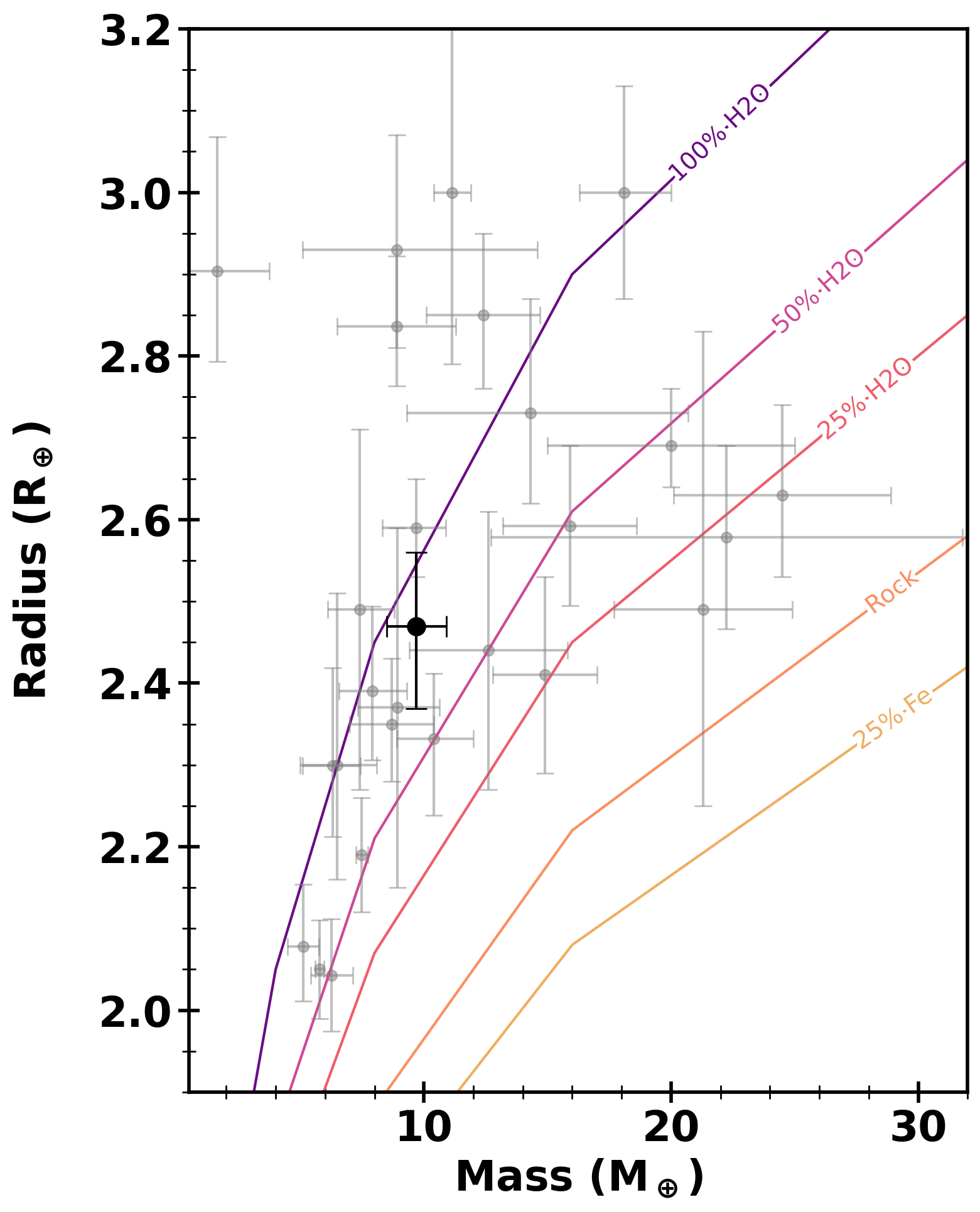}
    \caption{Mass-radius diagram for \ktwo sub-Neptunes ($R_P=2.0-3.0~R_\oplus$). The large black circle is K2-2 b, while the small gray circles are other sub-Neptunes with measured masses from the NASA Exoplanet Archive. The lines represent composition tracks from \cite{Zeng:2016}. }
    \label{fig:mvr}
\end{figure}

K2-2 b has a radius of $2.47^{+0.10}_{-0.09}~R_\oplus$ and a mass of $9.7\pm1.2~M_\oplus$. This yields a bulk density of $3.53^{+0.63}_{-0.57}$ g cm$^{-3}$, which is twice that of Neptune (1.638 g cm$^{-3}$). According to the composition models from \cite{Zeng:2016}, it is likely K2-2 b has a high water content (Figure \ref{fig:mvr}). While it is consistent with 100\% water, a more physically motivated solution would be a rocky core with an extended envelope of volatiles including a H/He envelope. More observations are needed to place further constraints on the planetary composition. 

The mass of K2-2 b was updated in a recent in-depth radial velocity study of Kepler and K2 systems \citep{Bonomo:2023} to refine planet masses and identify cold Jupiters in systems containing small planets. \cite{Bonomo:2023} refined important planetary parameters such as the period (to 9.0949 $\pm$ 0.0026 days) and mass (to 10.1 $^{+1.2}_{-1.1}$ $M_\oplus$), and did not find any long-term trends in the RVs that could correspond to a long-period companion. We used the same RV observations from this work (in addition to those from \citealt{Vanderburg:2015-K2-2}) but with improved precision from improved modeling of the stellar activity using the CALM technique (see Section \ref{sec:ccf}) in our global fit, and when combined with the other photometric and spectroscopic data, we were able to refine these measurements and uncover a potential outer companion due to a long-term trend in the RVs.

\subsection{RV trend}
\label{sec:rvtrendresults}

As mentioned in Section \ref{sec:RVs}, there is a long-term trend in the radial velocities (see Figure \ref{fig:rvs}) after correcting for stellar variability. To test the possibility of a second planet or star within the system, we reran the fit described in Section \ref{sec:GlobalFit} but allowed \texttt{EXOFASTv2} to fit for a second planet within the RVs only. We note that there is no additional transit signal detected in any photometric data sets used in this analysis. However, preliminary fits did not converge nor provide any useful constraint on the period of a potential companion, even with improved constraints on K2-2 b. Figure \ref{fig:rvs} shows the long-term trend in the RVs, and our resulting best-fit model from \texttt{EXOFASTv2}. It is clear that the period of this secondary companion is much longer term than the extent of our RV data set from HARPS-N ($\sim$2500 days). We instead model the long-term trend with a quadratic acceleration term. Our best-fit results find a linear slope in the RVs of $0.0024\pm0.0004$ m s$^{-1}$ with a quadratic term of $1.33E-06\pm3.6E-07$ m s$^{-1}$ day$^{-2}$ to best represent the long-term RV trend.

The observed RV trend may correspond to an additional companion to K2-2 with an orbital separation of several AU. \citet{Vanderburg:2015-K2-2} acquired high-resolution imaging observations of the star and did not detect any stellar companions between $0.1-5.0$" ($\approx6-310$~AU). This non-detection, combined with the relatively small amplitude of the RV acceleration, suggests that this outer companion could be a planet or a brown dwarf.

As K2-2 was observed by \textit{Hipparcos}, it is possible to place additional constraints on any outer companions using \textit{Hipparcos-Gaia} astrometry \citep{Brandt:2018, Brandt:2021}. If a massive companion exists at a separation of several AU from K2-2, it would likely generate a significant astrometric acceleration between \textit{Hipparcos} and \textit{Gaia}. However, no significant acceleration is detected in the \textit{Hipparcos-Gaia} astrometry, with $\chi^2=2.3$ for a constant proper motion \citep{Brandt:2021}. The astrometric precision for K2-2 is $\sim$0.07~mas~yr$^{-1}$, equivalent to $\sim$20~m~s$^{-1}$ at the $62.48\pm0.18$~pc distance of the system. This means that a net \textit{Hipparcos-Gaia} velocity change greater than $\gtrsim$100~m~s$^{-1}$ can be excluded at 5$\sigma$ confidence. This non-detection largely excludes the existence of massive companions ($\gtrsim$10~$M_J$) orbiting K2-2 within several AU. However, a planetary-mass companion could be reconciled with the astrometric non-detection.

Continued RV monitoring of the K2-2 system is needed to constrain the further evolution of the RV trend, providing some constraints on the fundamental parameters of the possible second planet in the system.



\subsection{Future work}
\label{sec:futurework}

The K2 mission was driven by the community, which led to planets orbiting much brighter host stars than the original Kepler mission, targets well suited for detailed characterization. Although characterization might be challenging with current facilities, K2-2 b is a worthwhile target for ongoing monitoring and targeted observations. Following the \citet{Kempton:2018} prescription for the transmission spectroscopy metric (TSM), we find that K2-2 b has a TSM of $50.0^{9.2}_{8.7}$, which falls just below the lowest value suggested for target prioritization for JWST. However, when compared to the other $\sim160$ sub-Neptunes ($R_P=2.0-3.0~R_\oplus$) in the K2 catalog, the TSM of K2-2 b is the fifth highest, suggesting that it is a suitable candidate for studying sub-Neptunes in closer detail. Monitoring the radial velocities of K2-2 would allow for more refined constraints on the stellar activity, and possibly uncover additional long-period and/or low-mass candidates in the system. 

The co-moving white dwarf (WD) companion to K2-2 provides an avenue to measure a precise age for the system if the mass and age for the WD can be determined. The stellar parameters were calculated as part of a catalog of all WDs within Gaia EDR3\footnote{Gaia EDR3 source\_id 2645940445519931520} by \cite{Gentile-Fusillo:2021}. The mass, effective temperature, and surface gravity were determined for three different atmospheric compositions: pure H, pure He, and a mix of H and He (see Table \ref{tab:wd}). Assuming the highest mass value from the models (pure-H, $0.52\pm0.04~M_\odot$), we find a lower limit on the cooling age of 1.13$\pm$0.13 Gyr. While this current age estimate does not constrain the system age further, more precise photometry and measuring the spectrum of the WD would constrain the mass (and system age) more reliably than Gaia photometry alone.

\begin{deluxetable}{lccc}[ht]
\tablecaption{Stellar parameters for the white dwarf companion of K2-2 from \cite{Gentile-Fusillo:2021}.}
\tabletypesize{\footnotesize}
\tablehead{\colhead{~~~Composition} & \colhead{$T_{\mathrm{eff}}$ (K)} & \colhead{$\log g$ (cgs)} &\colhead{Mass ($M_\odot$)}}
\startdata
~~~~H\dotfill &$7519\pm195$\dotfill & $7.88\pm0.08$ & $0.52\pm0.04$    \\
~~~~He\dotfill & $7395\pm189$\dotfill &  $7.82\pm0.06$  &  $0.47\pm0.02$    \\
~~~~H+He\dotfill & $7083\pm167$\dotfill & $7.71\pm0.07$ & $0.44\pm0.03$    \\
\enddata
\label{tab:wd}
 \begin{flushleft} 
  \footnotesize{
    }
\end{flushleft}
\end{deluxetable}

\section{Conclusion}
\label{sec:conclusion}
With thousands of exoplanets discovered to date, some will inevitably be ``lost" (unconstrained ephemerides) or forgotten as newer discoveries peak the interest of the community. Unfortunately, these lost planets may be excellent targets for detailed characterization with JWST \citep{Gardner:2006}, but are not accessible due to large uncertainties in future transit times. K2-2 b was the first planet discovered during the Two-Wheeled Concept Engineering Test of the K2 mission \citep{Howell:2014}, showing very quickly that K2 would be a successful repurposing of the Kepler spacecraft. By combining observations from multiple NASA missions along with key ground-based follow up that span nearly a decade, we have recovered the lost transit ephemeris of K2-2 b. In addition to being the first K2 planet, it is also well-suited for studying the atmosphere of a hot sub-Neptune as it orbits a bright ($K$$\sim$8.03) K-dwarf. This would be a valuable measurement since it sits on the high-mass peak of the sub-Neptune radius valley \citep{Owen:2012, Fulton:2017} and could provide insight to the formation and evolution of sub-Neptunes. Our updated ephemeris ($P=9.1004157^{+4.1E-06}_{-4.5E-06}$ days, $T_0 = 2458072.29291^{+0.00062}_{-0.00061}$ BJD) confirms the false detection from the MOST satellite \citep{Vanderburg:2015-K2-2} that led to a $\sim$40$\sigma$ offset to the true period. Systems like K2-2 show the importance of continued monitoring of exoplanet systems and dedicated ephemeris refinement efforts like the K2 \& TESS Synergy project \citep{Ikwut-Ukwa-SynergyI:2020, Thygesen:2023}, ExoClock \citep{Kokori-ExoclockI:2021,Kokori-ExoclockII:2022,Kokori-2023Exoclock3}, Exoplanet Watch \citep{Zellem:2019, Zellem:2020}, and ORBYTS \citep{Edwards:2019, Edwards:2020, Edwards:2021}. 

\software{Lightkurve \citep{Lightkurve:2018}, EXOFASTv2 \citep{Eastman:2013, Eastman:2019}, AstroImageJ \citep{Collins:2017}}
\facilities{TESS, K2, Spitzer, MEarth, University
of Louisville Manner Telescope (ULMT), Telescopio Nazionale Gailieo 3.58 m (HARPS-N), FLWO 1.5m (Tillinghast Reflector Echelle Spectrograph; TRES), Gaia, MAST.}

\acknowledgements
We thank the referee for the feedback that greatly improved the manuscript. ET and JER acknowledge support for this project from NASA'S TESS Guest Investigator program (G04205, P.I. Rodriguez). We thank Jason Eastman for the lengthy discussions on the inner workings of {\tt EXOFASTv2}. ET would like to thank the Quad Fellowship for support. Z.L.D. would like to thank the generous support of the MIT Presidential Fellowship and to acknowledge that this material is based upon work supported by the National Science Foundation Graduate Research Fellowship under grant No. 1745302. Z.L.D would like to acknowlegde the MIT Collamore-Rogers Fellowship. 

The MEarth Team gratefully acknowledges funding from the David and Lucile Packard Fellowship for Science and Engineering (awarded to D.C.). This material is based upon work supported by the National Science Foundation under grants AST-0807690, AST-1109468, AST-1004488 (Alan T. Waterman Award), and AST-1616624, and upon work supported by the National Aeronautics and Space Administration under Grant No. 80NSSC18K0476 issued through the XRP Program. This work is made possible by a grant from them John Templeton Foundation. The opinions expressed in this publication are those of the authors and do not necessarily reflect the views of the John Templeton Foundation.

This research has made use of SAO/NASA's Astrophysics Data System Bibliographic Services. This research has made use of the SIMBAD database, operated at CDS, Strasbourg, France. This work has made use of data from the European Space Agency (ESA) mission {\it Gaia} (\url{https://www.cosmos.esa.int/gaia}), processed by the {\it Gaia} Data Processing and Analysis Consortium (DPAC, \url{https://www.cosmos.esa.int/web/gaia/dpac/consortium}). Funding for the DPAC has been provided by national institutions, in particular the institutions participating in the {\it Gaia} Multilateral Agreement. This work makes use of observations from the LCO network. This work is based [in part] on observations made with the Spitzer Space Telescope, which was operated by the Jet Propulsion Laboratory, California Institute of Technology under a contract with NASA. This research made use of Lightkurve, a Python package for \textit{Kepler} and \TESS\ data analysis. The data presented in this paper were obtained from the Mikulski Archive for Space Telescopes (MAST) at the Space Telescope Science Institute. Data from TESS Sectors 42 and 70 can be accessed at \dataset[doi:10.17909/t9-nmc8-f686]{https://doi.org/10.17909/t9-nmc8-f686} and K2 Campaign 0 at \dataset[doi:10.17909/T9F88F]{https://doi.org/10.17909/T9F88F}. Some data in this work were accessed at ExoFOP, accessible at \dataset[doi:10.26134/ExoFOP3]{https://doi.org/10.26134/ExoFOP3}. The Spitzer data used in this work can be found at \dataset[doi:10.26131/IRSA430]{https://doi.org/10.26131/IRSA430}.

Funding for the \TESS\ mission is provided by NASA's Science Mission directorate. We acknowledge the use of public \TESS\ Alert data from pipelines at the \TESS\ Science Office and at the \TESS\ Science Processing Operations Center. This research has made use of the Exoplanet Follow-up Observation Program website, which is operated by the California Institute of Technology, under contract with the National Aeronautics and Space Administration under the Exoplanet Exploration Program. Resources supporting this work were provided by the NASA High-End Computing (HEC) Program through the NASA Advanced Supercomputing (NAS) Division at Ames Research Center for the production of the SPOC data products.

\clearpage
\bibliographystyle{apj}
\bibliography{refs}



\end{document}